\PassOptionsToPackage{dvipsnames}{xcolor}

\documentclass[ACS,STIX1COL]{WileyNJD-v2}
\usepackage{moreverb}
\usepackage{graphicx} 
\definecolor{MyGreen}{RGB}{141,208,138}
\definecolor{MyBlue}{RGB}{134,190,220}
\definecolor{MyOrange}{RGB}{252,128,96}
\definecolor{MyPurple}{RGB}{174,173,211}

\usepackage{amsmath}
\usepackage{multirow}
\usepackage{placeins}
\usepackage{subcaption}
\articletype{article}

\graphicspath{ {./Figures/} }

\begin{document}

\title{A comparative study of analytical models of diffuse reflectance in homogeneous biological tissues: Gelatin based phantoms and Monte Carlo experiments.}
\author[1]{Anisha Bahl*}
\author[1]{Silvere Segaud}
\author[1]{Yijing Xie}
\author[1,3]{Jonathan Shapey}
\author[2]{Mads Bergholt}
\author[1]{Tom Vercauteren}

\address[1]{\orgdiv{School of Biomedical Engineering \& Imaging Sciences}, 
\orgname{King's College London}, 
\orgaddress{1 Lambeth Palace Road, London, United Kingdom}}
\address[2]{\orgdiv{Department of Craniofacial Development and Stem Cell Biology}, 
\orgname{King's College London}, 
\orgaddress{Guy's Tower, Great Maze Pond, London, United Kingdom}}
\address[3]{\orgname{King's College Hospital}, 
\orgaddress{Denmark Hill, London, United Kingdom}}

\corres{*Anisha Bahl, School of Biomedical Engineering \& Imaging Sciences, King's College London, 1 Lambeth Palace Road, London, United Kingdom. \email{anisha.bahl@kcl.ac.uk}}

\abstract[Abstract]{
Information about tissue oxygen saturation ($StO_2$) and other related important physiological parameters can be extracted from diffuse reflectance spectra measured through non-contact imaging.
Three analytical optical reflectance models for homogeneous, semi-infinite, tissue have been proposed (Modified Beer-Lambert, Jacques 1999, Yudovsky 2009) but these have not been directly compared for tissue parameter extraction purposes. 
We compare these analytical models using Monte Carlo simulated diffuse reflectance spectra and 
controlled gelatin-based phantoms
with
measured diffuse reflectance spectra and known ground truth composition parameters.
The Yudovsky model performed best against Monte Carlo simulations and measured spectra of tissue phantoms in terms of goodness of fit and parameter extraction accuracy followed closely by Jacques' model. 
In this study, Yudovsky's model appeared most robust, however our results demonstrated that both Yudovsky and Jacques models are suitable for modelling tissue that can be approximated as a single, homogeneous, semi-infinite slab.}

\keywords{Biological models; Oxygen saturation; Monte Carlo simulations; Imaging phantoms; Gelatin}

\maketitle 

\section{Introduction}\label{sec:intro}
Tissue oxygen saturation ($StO_2$) is a key metric that could be used in surgery to determine tissue viability, tumour localisation, and vessel flow \cite{Takami2017, Hughes2019, Richardson2023}.
There are, however, no currently established, spatially-resolved, intra-operative methods of quantitatively determining $StO_2$.
In many clinical applications indocyanine green fluorescence angiography (ICG) is used to display perfusion as 
blood flow through vessels can be visualised by the fluorescence of the indocyanine dye given to the patient \cite{Hackethal2018}.
Whilst this method has shown clinical benefits \cite{Renna2023, Teng2021}, it cannot quantify the $StO_2$ of the tissues being visualised leading to subjective interpretation.
There has been increasing promise that diffuse reflectance spectroscopic techniques could be used to obtain $StO_2$ intra-operatively. Hyperspectral Imaging (HSI) is one such technique that obtains a diffuse reflectance spectrum for each pixel of an image \cite{Kulcke2018, Taylor-Williams2022}, enabling intra-operative, non-contact, spatially-resolved $StO_2$ extraction without the need for a contrast agent. 

There have been many proposed methods to extract $StO_2$ parameters from tissue diffuse reflectance spectra ranging from the simplest ratiometric two-wavelength methods to more sophisticated analytical models \cite{MacKenzie2018} or deep-learning based approaches \cite{Ayala2023}. Using two or three wavelength models can provide good results but require data to be captured at these precise wavelengths placing large constraints on the measurement devices used and significant model assumptions. Analytical models can be applied to ranges of wavelengths for which they are developed and so these are favoured in this work. Tissue models tend to either be based on modifications of the Beer-Lambert model, or on the diffusion approximation particularly using the Kulbelka-Munk approximation method\cite{MacKenzie2018}. Three models that can be applied to homogeneous semi-infinite tissue in the visible wavelength range (450-650nm) include 1) a modified Beer-Lambert model \cite{Clancy2015}; 2) a more elaborate Beer-Lambert-based model proposed by Jacques \cite{Jacques1999}; and 3) a semi-empirical model based on the Kulbelka-Munk theory as described by Yudovsky \cite{Yudovsky2009}.
Alternatively, Monte Carlo methods may be used to model tissues for a range of wavelengths. Whilst Monte Carlo is an established approach for modelling light-tissue interaction, it is computationally challenging to apply in an inverse problem setting making the estimation of optical properties from tissue spectra non trivial on the basis of Monte Carlo simulations alone.
To address the need for improved inverse modelling utilising Monte Carlo, Inverse Adding Doubling (IAD)\cite{Prahld} has been proposed, however this is also slow iterative process and therefore not as efficient as approaches based on analytical models.
A limitation of IAD is that, although a single reflectance measurement can be used to analyse semi-infinite samples with IAD, its primary focus is the extraction of optical properties from slabs of tissue with known thickness using both transmission and reflection measurements.
In this work, IAD is used to provide information on optical properties from tissue phantom slabs but is not assessed in the same capacity as the other 
analytical models listed above. 

Since there is no standard method of determining spatially resolved, ground truth, optical properties of tissues, 
simulations and physical
phantoms must be used
for testing purposes.
Gelatin-based phantoms are commonly used to allow tunable absorption and scattering properties in solid phantoms \cite{Pogue2006, Gautam2023}.
In this work we compare three major analytical models against both Monte Carlo simulated spectra, and measured spectra from gelatin-based tissue phantoms with known ground-truth constituent quantities. These datasets cover a large parameter range mimicking that expected in biological tissue\cite{Jacques2013}. Gelatin phantoms are constructed to optically mimic biological tissue absorption and scattering ranges as closely as possible, whilst maintaining well-defined optical properties with known ground truth. We present a first direct comparison in the performance of these models against simulated and measured spectra, and aim to evaluate their potential for accurate $StO_2$ extraction. This evaluation is quantified in terms of the forward models (utilising ground truth values within the models to predict the diffuse reflectance spectra), and for the inverse problem solving (evaluating the fidelity of the parameter extraction when fitting to the simulated or measured data).
Our results provide a comparison which can be used to inform the translation of these models into clinical applications.

\section{Materials and methods}\label{sec:methods}
\subsection{Tissue models}\label{sec:methodtissuemodels}
Three analytical models are compared and evaluated in this work: modified Beer-Lambert \cite{Clancy2015}, Jacques 1999 \cite{Jacques1999}, and Yudovsky 2009 \cite{Yudovsky2009}. Due to the absorbing and scattering nature of biological tissue these models are used to analyse diffuse reflectance spectra which result from propagation of light via many optical paths through the tissue.

\subsubsection{Common absorption and scattering model}
All three forward models use the wavelength dependent absorption and reduced scattering coefficients ($\mu_a(\lambda)$ and $\mu_s'(\lambda)$) of the tissue as inputs
to compute a
diffuse reflectance spectrum. 
The reduced scattering coefficient comprises of the scattering coefficient ($\mu_s(\lambda)$) and tissue anisotropy ($g$) as follows: 
\begin{equation}
    \mu_s'(\lambda) = \mu_s(\lambda) \times (1-g)
    \label{eq:reducedscattering}
\end{equation}
For biological tissue, the reduced scattering coefficient can be well approximated using Mie theory \cite{Jacques2013}: 
\begin{equation}
    \mu_s'(\lambda) = a(\frac{\lambda}{500})^{-b}
    \label{eq:Mie}
\end{equation}
where $a$ and $b$ are Mie scattering coefficients that range between 8\textrm{$cm^{-1}$} and 70\textrm{$cm^{-1}$}, and 0.1 and 3.3 respectively depending on the tissue microstructure \cite{Jacques2013}. 
The absorption coefficient of most internal, homogeneous, single-layer tissues is dominated by haemoglobin in the visible region (450-650 nm)\cite{JacquesAbs} and can be modelled using the following equation \cite{Yudovsky2009}: 
\begin{equation}
\begin{aligned}
    & \mu_a(\lambda) = f_{blood}\mu_{a, blood}(\lambda) + (1 - f_{blood})\mu_{a, back}(\lambda) \\
    & \textrm{where} \\
    & \mu_{a, blood}(\lambda) = c_{HbT}\frac{\ln(10)}{64500}[StO_2 \epsilon_{HbO_2}(\lambda) + (1 - StO_2)\epsilon_{Hb}(\lambda)] \\
    & \textrm{and} \\
    & \mu_{a, back}(\lambda) = 7.84\times10^8 \lambda^{-3.255}
\end{aligned}
\label{eq:mua}
\end{equation}
Here $StO_2$ refers to oxygen saturation, $c_{HbT}$ refers to the total concentration of haemoglobin in whole blood commonly taken as 150\textrm{$gL^{-1}$}\cite{Prahlb}, with $StO_2$ denoting the fraction of this that is oxygenated ($HbO_2$) and the remainder is deoxygenated ($Hb$), and ranges between 0-100\%\cite{Yudovsky2009}. $f_{blood}$ is the volume fraction of tissue occupied by blood which has absorption coefficient $\mu_{a, blood}(\lambda)$ and is examined in the range of 0.2-7\%\cite{Yudovsky2009}, with the remainder of tissue having background absorption $\mu_{a, back}(\lambda)$\cite{Yudovsky2009}.
Finally, $\epsilon_{HbO_2}(\lambda)$ and $\epsilon_{Hb}(\lambda)$ denotes the wavelength-dependent extinction coefficients of the chromophores $HbO_2$ and $Hb$ which are found in the literature\cite{Prahlb}. 

\subsubsection{Modified Beer-Lambert}
The modified Beer-Lambert utilises these inputs to model absorption ($A(\lambda)$) and diffuse reflectance ($R(\lambda)$) as follows: 
\begin{equation}
\begin{aligned}
    & A(\lambda) = L\mu_a(\lambda) + \mu_s'(\lambda) \\
    & R(\lambda) = \exp{\left(-\frac{A(\lambda)}{100}\right)}
\end{aligned}
\label{eq:modBL1}
\end{equation}
Conventionally, $\mu_a(\lambda)$ and $\mu_s'(\lambda)$ are quoted in units of cm\textrm{$^{-1}$}, however in this model units of mm\textrm{$^{-1}$} are required for diffuse reflectance units of \% hence a factor of 100 is included. $L$ describes a differential path length to account for the variety of photon path lengths through scattering media. $L$ is often simplified to be equal to 1 \cite{Clancy2015} and $\mu_s'(\lambda)$ is often modelled as a wavelength-independent constant\cite{Clancy2015, Ma2016}. 
To allow for further flexibility in this model,
we introduce linear scaling hyperparameters ($M_{1-3}$), as shown in Equation \eqref{eq:modBL2}, that can be fitted to Monte Carlo simulations at the refractive index of interest. 
\begin{equation}
\begin{aligned}
    & A(\lambda) = M_1\mu_a(\lambda) + M_2\mu_s'(\lambda) + M_3 \\
    & R(\lambda) = \exp{\left(-\frac{A(\lambda)}{100}\right)}
\end{aligned}
\label{eq:modBL2}
\end{equation}

\subsubsection{Jacques 1999}
The Jacques model is also based on the Beer-Lambert model, where an assumption is made that the ensemble of path lengths experienced by photons in a tissue can be approximated by a single wavelength-dependent path length $L(\lambda) = A(\lambda)\delta(\lambda)$ where $A(\lambda)$ and $\delta(\lambda)$ are defined in Equation \ref{eq:Jacques}. This results in the following model with the hyperparameters ($M_{1-3}$) which the authors
fit
to Adding Doubling simulations \cite{Jacques1999}. In our work, we refit these to Monte Carlo simulations since these are considered the gold standard optical simulation method and improve the model fitting results. 
\begin{equation}
\begin{aligned}
    & N'(\lambda) = \frac{\mu_s'(\lambda)}{\mu_a(\lambda)} \\
    & \delta(\lambda) = \frac{1}{\sqrt{3\mu_a(\lambda)[\mu_a(\lambda) + \mu_s'(\lambda)]}} \\
    & A(\lambda) = M_1 + M_2\exp \left[ \frac{\ln(N'(\lambda))}{M_3} \right] \\
    & R(\lambda) = \exp[-A(\lambda)\delta(\lambda)\mu_a(\lambda)] \\
\end{aligned}
\label{eq:Jacques}
\end{equation}

\subsubsection{Yudovsky 2009}
The original Yudovsky model presents a complex analytical derivation \cite{Yudovsky2009} with a simplified formulation subsequently presented in their Erratum \cite{Yudovsky2015}. The model takes the reduced albedo ($w'(\lambda)$) as input and returns the diffuse reflectance spectra ($R(\lambda)$). This provides an easily applicable model with hyperparameters ($M_{1-6}$) which are quoted for a refractive index ($n$) of 1.44. We found that these can be fitted to Monte Carlo spectra to allow for use with other refractive indices. 
\begin{equation}
\begin{aligned}
    & w'(\lambda) = \frac{\mu_s'(\lambda)}{\mu_a(\lambda) + \mu_s'(\lambda)} \\
    & R = M_1 + M_2\exp{\left[ M_3w'(\lambda)^{M_4}\right]} + \frac{M_5}{1.02 - M_6} \\
\end{aligned}
\label{eq:Yudovskysingle}
\end{equation} 

\subsubsection{Fitting model hyperparameters}
All analytical models considered use hyperparameters to account for the media's refractive index. A Monte Carlo dataset is generated for each refractive index as in Section \ref{sec:methodsMC}. To fit the hyperparameters, the ground truth tissue parameters are inputted into an analytical model for each spectrum in the dataset and a single set of hyperparameters are fitted using a non-linear least squares fitting approach.

\subsection{Generating reference datasets of diffuse reflectance spectra}\label{sec:methodreference}
In this work, we aim to compare the three models discussed in Section \ref{sec:methodtissuemodels} against datasets with known ground truth. These datasets of diffuse reflectance spectra are generated either by simulation using Monte Carlo (Section \ref{sec:methodsMC}) or measurement of controlled, gelatin-based, tissue phantoms (Section \ref{sec:methodsphantoms}).

\subsubsection{Monte Carlo}\label{sec:methodsMC}
The models take $\mu_a(\lambda)$ and $\mu_s'(\lambda)$ as inputs. These are given for bulk tissue in Equations \eqref{eq:mua} and \eqref{eq:Mie} and used to generate Monte Carlo simulated reference spectra. Monte Carlo takes $\mu_s(\lambda)$ as input, not $\mu_s'(\lambda)$, so a random value of anisotropy ($g$) is chosen per spectrum between 0.7-0.9\cite{Yudovsky2009} and used in Equation \eqref{eq:reducedscattering} for conversion to $\mu_s(\lambda)$. The variable parameters are $a$, $b$, $StO_2$, and $f_{blood}$ which are bounded to 8-70cm\textrm{$^{-1}$}, 0.1-3.3, 0-100\%, 0.2-7\% respectively to mimic biological tissue\cite{Yudovsky2009, Jacques2013}; a value of each of these variables is randomly selected within these bounds to generate each of the 100 simulated spectra, meaning 100 values of each parameter are sampled. When fitting the inverse models to these spectra for parameter extraction (as described in Section \ref{sec:methodevaluate}), the same bounds are imposed on the fitting routine. These simulated spectra are generated using CUDAMCML\cite{Alerstam2008}, which is a GPU accelerated adaptation of the well-established MCML programme\cite{Wang1995}. It has been shown that the semi-infinite approximation is valid for thicknesses of above 1cm\cite{Zhang2014} so 100 spectra were simulated for each refractive index of 1.33, 1.35, and 1.44, by propagating 100,000 photons through the semi-infinite slabs approximated by a thickness of 3cm. The refractive indices were chosen to represent the common phantom and tissue use-cases of these models: 1.33 for use with water-based phantoms, 1.35 for use with gelatin-based phantoms (as in this work), and 1.44 for use with biological tissue.

\subsubsection{Controlled gelatin-based tissue phantoms}\label{sec:methodsphantoms}
Here we construct controlled, optical tissue phantoms with well-characterised components. By measuring these phantoms we constructed a dataset of measured spectra with well-defined ground truth which can be modelled using the analytical models.

\paragraph{Phantom composition and synthesis}\label{sec:methodsphantomcomposition}

\begin{figure}[htb]
    \centering
    \begin{subfigure}{0.6\textwidth}
        \includegraphics[width=\textwidth]{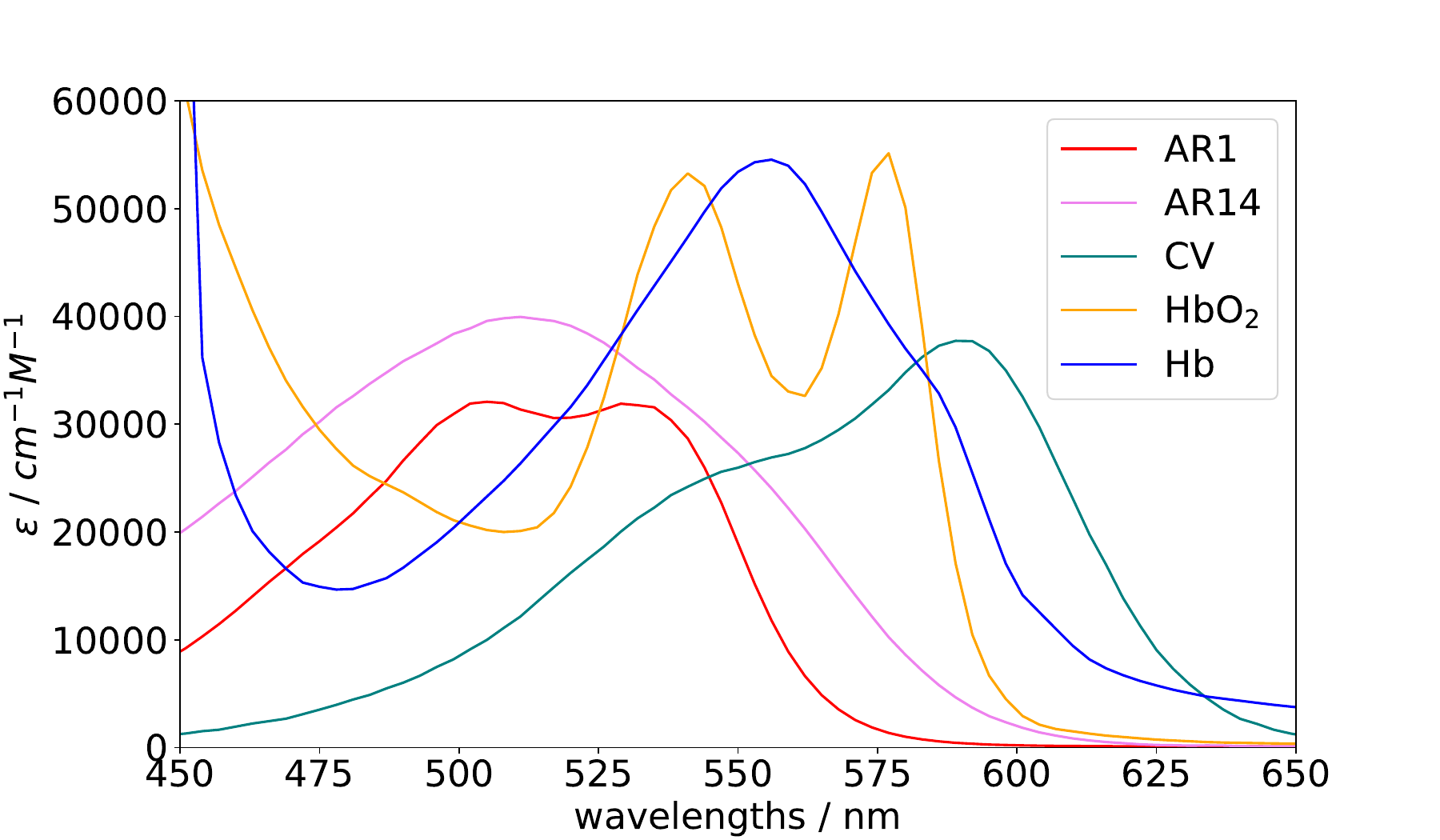}
        \caption{}
        \label{fig:exteff}
    \end{subfigure}
    \hfill
    \begin{subfigure}{0.44\textwidth}
        \includegraphics[width=\textwidth]{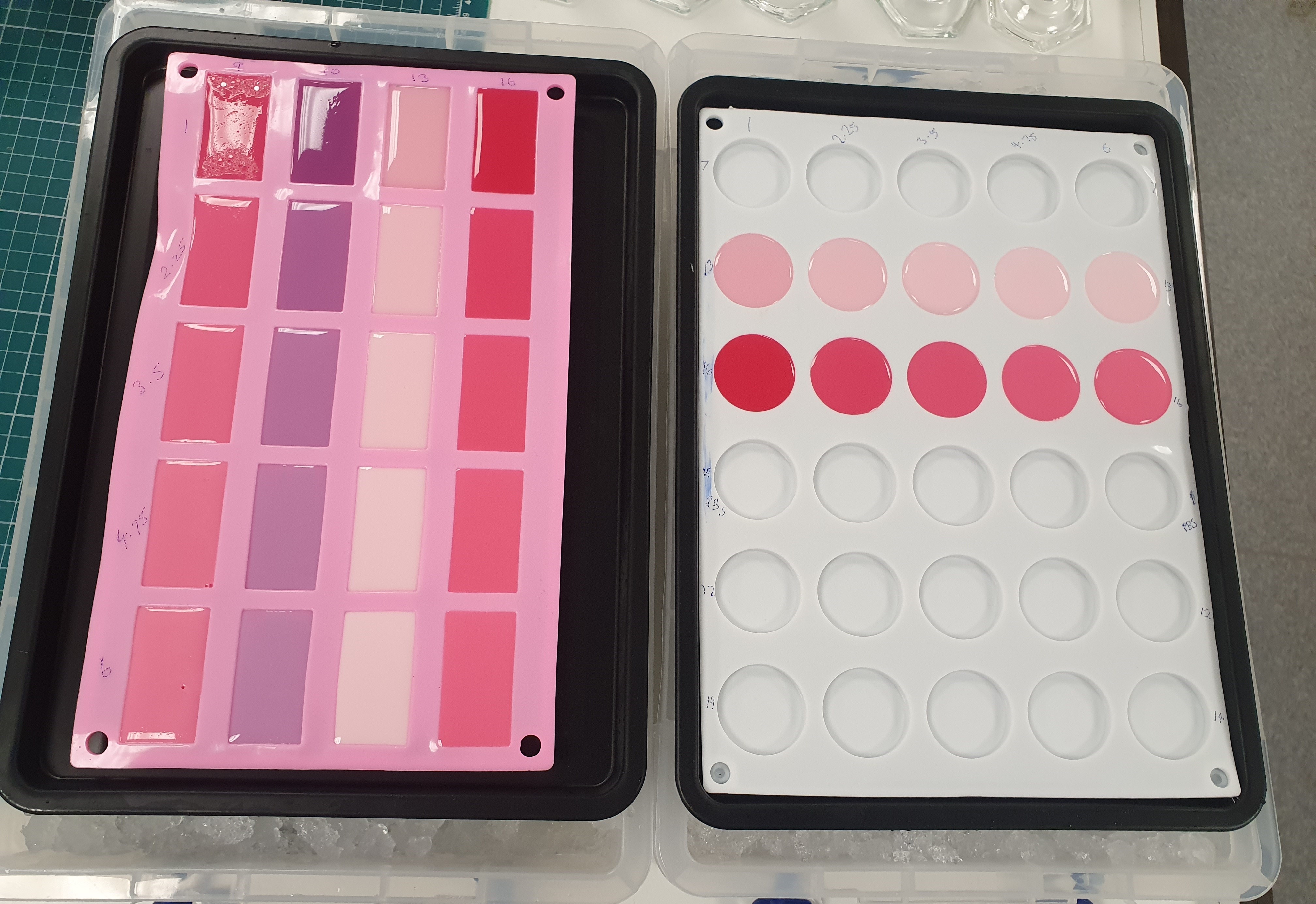}
        \caption{}
        \label{fig:icebath}
    \end{subfigure}
    \begin{subfigure}{0.245\textwidth}
        \includegraphics[width=\textwidth]{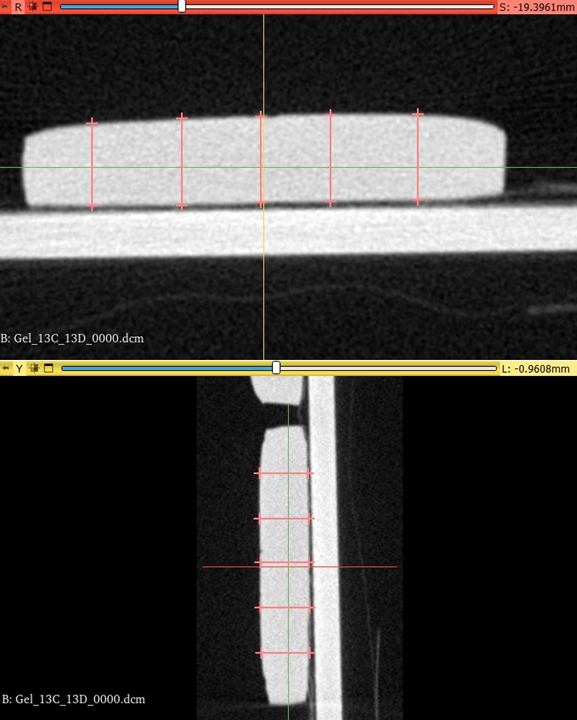}
        \caption{}
        \label{fig:DICOM_Eg}
    \end{subfigure}
    \caption{Figure regarding phantom development. Effective extinction coefficients of acid red 1 (AR1), acid red 14 (AR14), and crystal violet (CV) displayed with the extinction coefficients of oxygenated (HbO$_2$) and deoxygenated haemoglobin (Hb) (\ref{fig:exteff}). Image of 1cm depth (left) and 5mm depth (right) phantoms setting in ice bath (\ref{fig:icebath}). An example of a CT scan of a tissue phantom and 10 digital measurements taken of the phantom thickness calculated to be 6.11mm ($\pm$0.11mm) (\ref{fig:DICOM_Eg}).}
    \label{fig:phantommethods}
\end{figure}

The dyes acid red 1 (AR1, 210633, Merck, Germany) and acid red 14 (AR14, B22328, Fisher Scientific, England) are chosen to mimic the extinction coefficients of oxygenated and deoxygenated haemoglobin respectively, with a third dye of crystal violet (CV, C6158, Merck, Germany) chosen to investigate the effect of including further chromophores. As in the modelled tissue $\mu_a(\lambda)$ (Equation \eqref{eq:mua}), the total dye concentration is modelled independently of the relative ratios of each dye. To ensure the absorbance impact of each dye is approximately equal, a factor of $\frac{5}{3}$ is included for AR14 and $\frac{1}{2}$ for CV computationally by combining with the extinction coefficients to create effective extinction coefficients ($\epsilon_{eff}$) that have approximately equal impact. To reflect this experimentally, the concentrations of these dyes are modified by these factors. These effective extinction coefficients calculated from literature values\cite{PhotochemCAD}, alongside those of haemoglobin\cite{Prahlc}, can be seen in Figure \ref{fig:exteff}. 

The tissue phantoms are constructed with three overall dye concentrations corresponding to fractions of blood of 0.344\%, 3.44\%, and 6.88\% ie. $8\times10^{-6}$ moldm$^{-3}$, $8\times10^{-5}$ moldm$^{-3}$, and $1.6\times10^{-4}$ moldm$^{-3}$. This can be described as a 1x, 10x, and 20x concentration with a scale factor of $8\times10^{-6}$.
Two-dye configurations are constructed to investigate a range of AR1:AR14 ratios with some 3-dye configurations added. 

Intralipid is used to modulate the scattering coefficient of these phantoms which can be modelled with a Mie scattering function (Equation \eqref{eq:Mie}) using parameters within the range of tissues. 5 volume fractions of intralipid are chosen between 1\% and 6\% to cover a range of scattering parameters within those seen in biological tissue \cite{Jacques2013}. 

Each phantom consists of 6\% gelatin (Type A approximately 175g bloom, G2625, Merck, Germany) by mass 
and 0.5\% of 4\% formaldehyde (J60401, Fisher Scientific, England) to increase the melting point of the phantoms and allow their use at room temperature \cite{Pogue2006}. The remainder of each phantom consists of the dye solutions in a variety of ratios combined with intralipid at a variety of concentrations. 

Phantoms are constructed with the ratios given in Table \ref{tb:phantomratios}, where each configuration is constructed with each intralipid concentration. 

\begin{table}[ht!]
    \centering
    \caption{Table displaying the ratios of dyes [acid red 1 (AR1), acid red 14 (AR14), crystal violet (CV)] a used for each dye configuration alongside the total dye scaled concentration in arbitrary units.}
    \begin{tabular}{|c|c|c|c|}
        \hline
        Total dye scaled concentration & \multicolumn{3}{|c|}{Dye ratio} \\
        (arbitrary units) & AR1 & AR14 & CV \\
        \hline
        1 & 1 & 1 & 0 \\
        1 & 2 & 1 & 1 \\
        1 & 1 & 1 & 2 \\
        10 & 1 & 0 & 0 \\
        10 & 3 & 1 & 0 \\
        10 & 1 & 1 & 0 \\
        10 & 1 & 3 & 0 \\
        10 & 0 & 1 & 0 \\
        10 & 1 & 2 & 1 \\
        10 & 2 & 1 & 1 \\
        10 & 1 & 1 & 2 \\
        20 & 1 & 1 & 0 \\
        20 & 2 & 1 & 1 \\
        20 & 1 & 1 & 2 \\
        \hline
    \end{tabular}
    \label{tb:phantomratios}
\end{table}

\label{sec:methodsphantomsynthesis}
For each dye, stock solutions are made at double the required concentrations: 2x, 20x, and 40x in phosphate buffered saline (PBS, P4417, Merck, Germany) to allow for 1:1 dilution with intralipid as the final step. The stock concentrations are multiplied by a factor of $\frac{5}{3}$ for Acid Red 14 and $\frac{1}{2}$ for Crystal Violet to ensure similar absorbance impact. For each intralipid concentration, the stock solutions are made at double the required volume fractions: 2\% - 12\% to allow for 1:1 dilution with the dye solutions as the final step. The dye solutions combined in the correct ratios are heated to 45-50$^o$C with 12\% gelatin (i.e. double the required amount) until solvation. The solution is cooled to below 40$^o$C where 1\% formaldehyde (ie. double the quantity) is added and the solution is combined in equal quantities with the solution with double the desired intralipid concentration. This final solution now has the intended concentration of all constituents and is poured into a silicon mould in an ice bath, as seen in Figure \ref{fig:icebath}, to ensure homogeneity in the final phantom by rapidly setting prior to any density separation. Once cooled to below 10$^o$C these are placed in a fridge at 4$^o$C to fully set for at least 7 days before measurement. 

Each dye solution is also combined with gelatin with a 0\% intralipid solution for absorbance measurements.
Additionally, each intralipid solution is combined with a gelatin solution with no dyes to allow analysis of scattering due to intralipid concentration.

Moulds with a depth of 1 cm are used to allow measurement of diffuse reflectance within a semi-infinite regime\cite{Zhang2014}.
Monte Carlo simulations were used to confirm that this thickness was sufficient to be within the semi-infinite regime.
Some additional phantoms are constructed with a depth of 5 mm to allow for sufficient transmission for total transmittance measurements required for inverse adding doubling (IAD) analysis \cite{Prahld}. 

\begin{figure}[t!]
    \centering
    \begin{subfigure}{0.47\textwidth}
        \includegraphics[width=\textwidth]{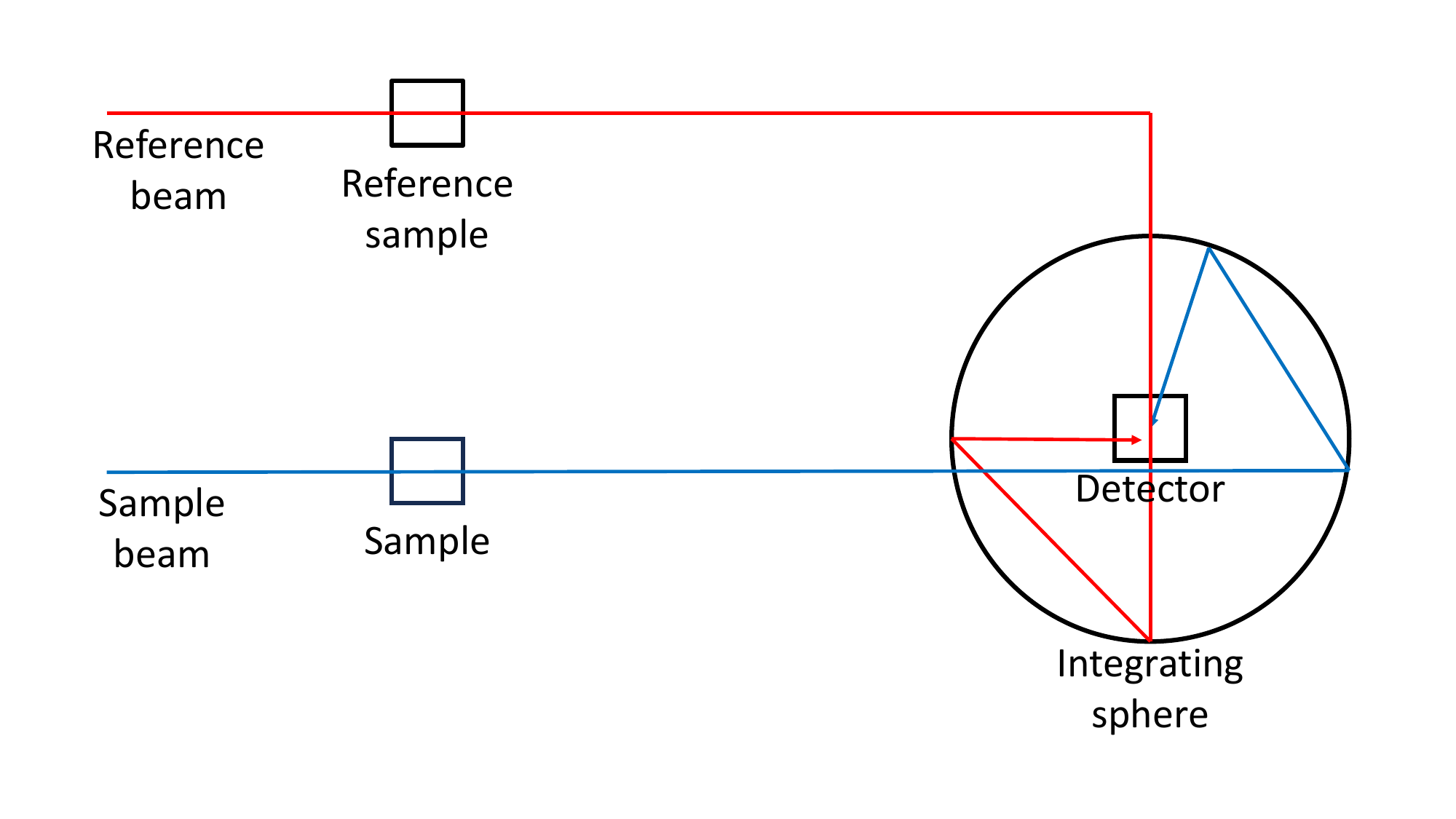}
        \caption{}
        \label{fig:spectrophotometer_abs}
    \end{subfigure}
    \begin{subfigure}{0.47\textwidth}
        \includegraphics[width=\textwidth]{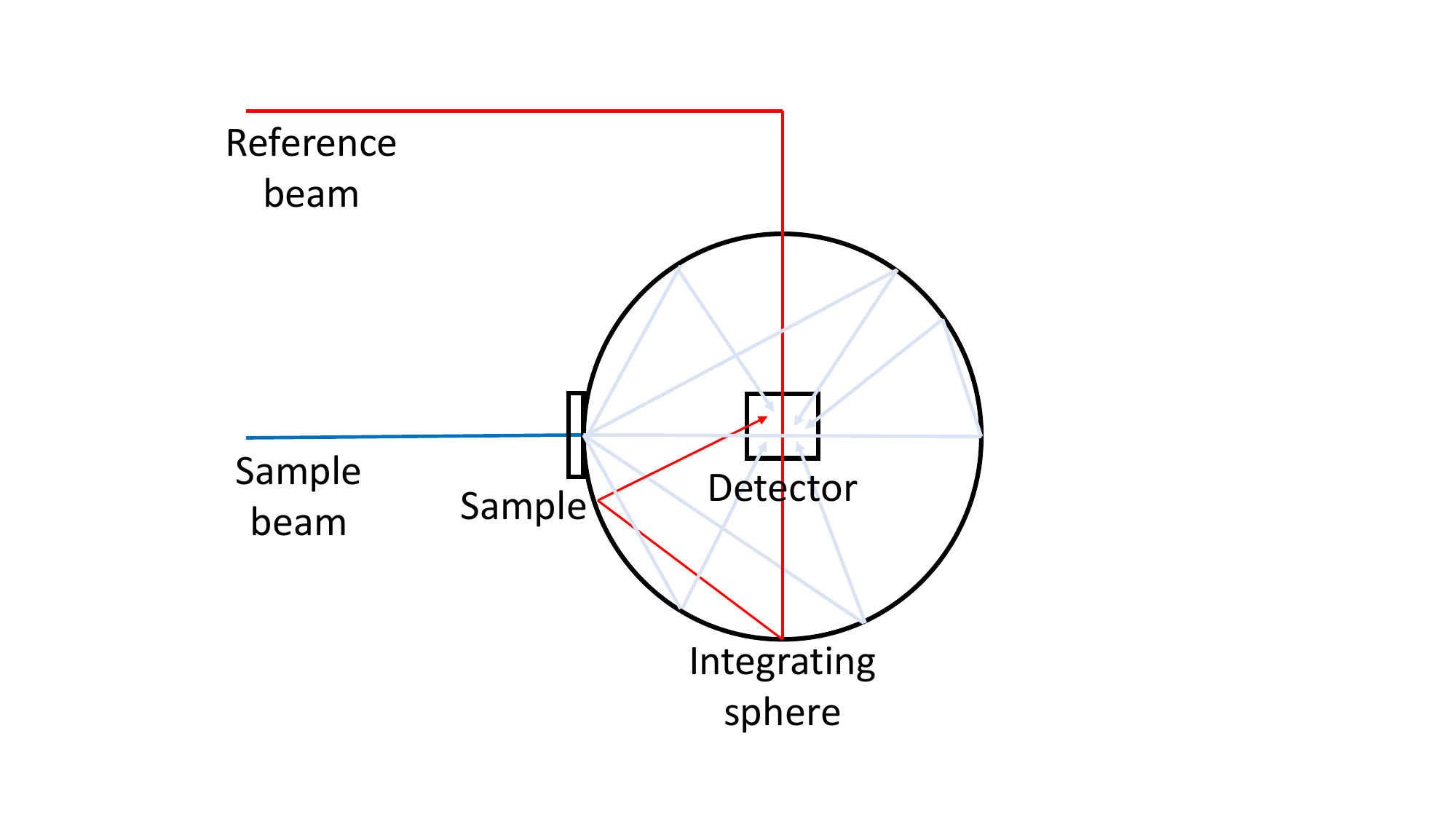}
        \caption{}
        \label{fig:spectrophotometer_dT}
    \end{subfigure}
    \hfill
    \begin{subfigure}{0.47\textwidth}
        \includegraphics[width=\textwidth]{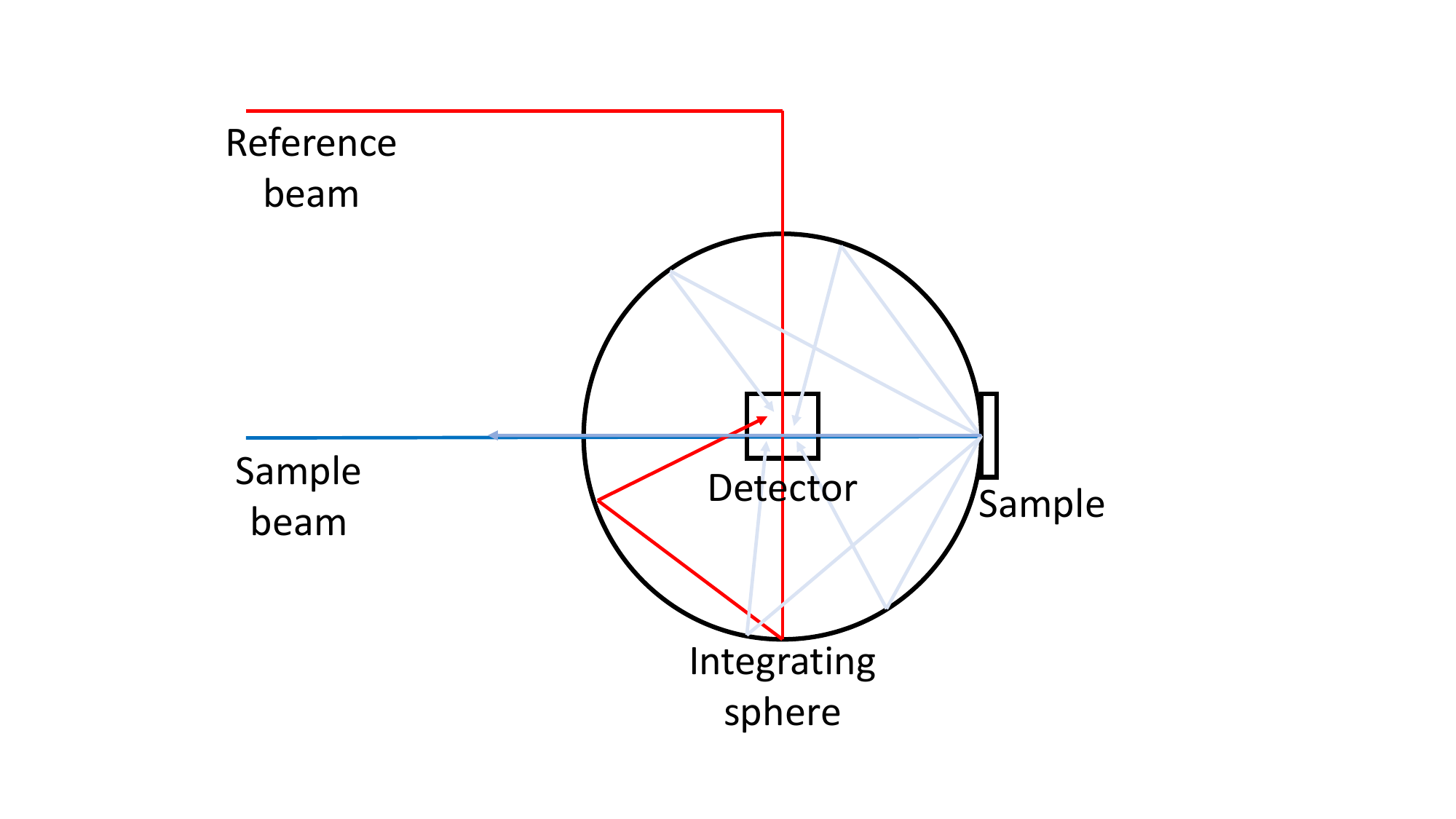}
        \caption{}
        \label{fig:spectrophotometer_dR}
    \end{subfigure}
    \begin{subfigure}{0.47\textwidth}
        \includegraphics[width=\textwidth]{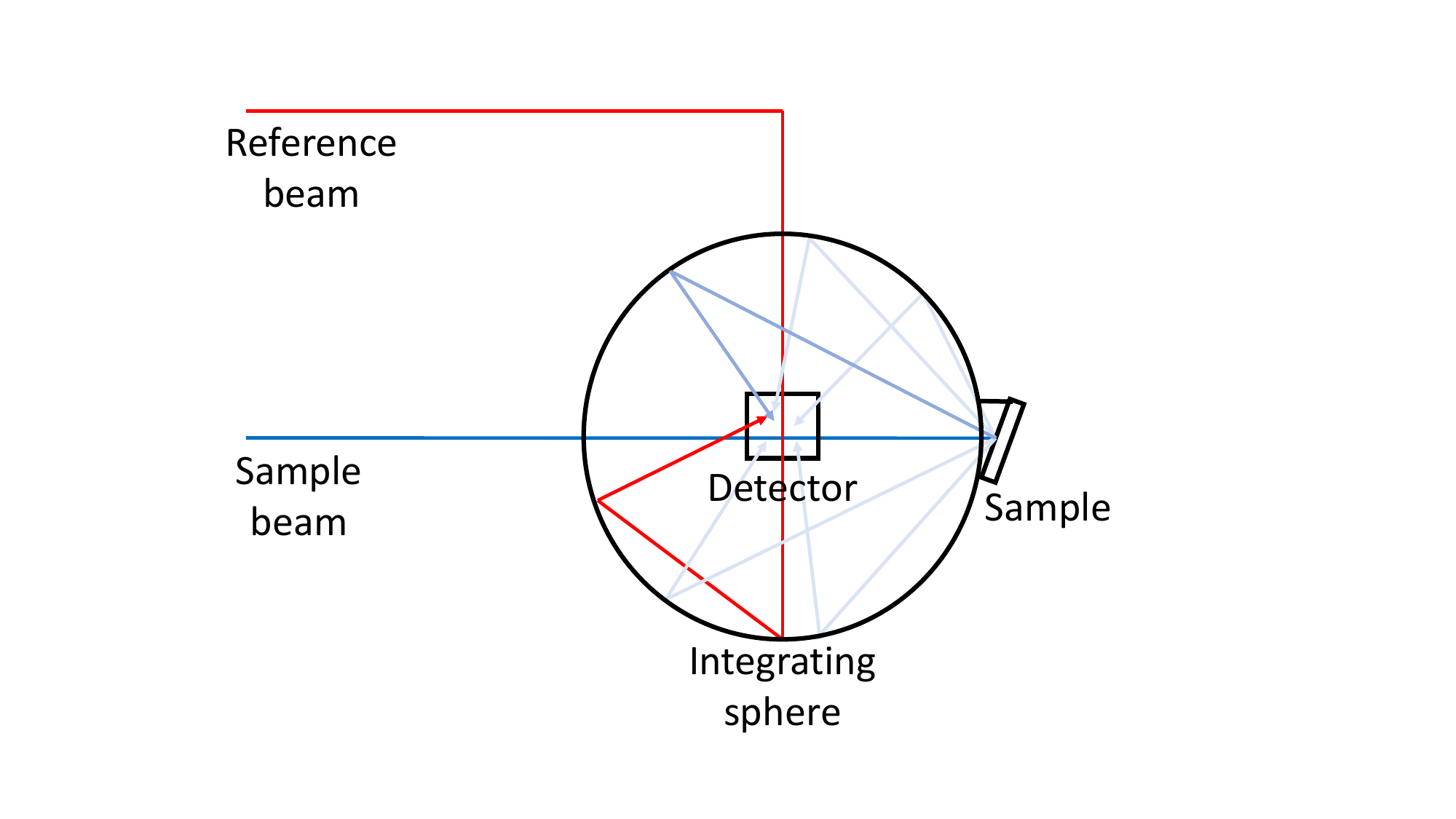}
        \caption{}
        \label{fig:spectrophotometer_tR}
    \end{subfigure}
    \caption{Depictions of measurement set-up for spectrophotometer measurements of gelatin based tissue phantoms of absorbance (\ref{fig:spectrophotometer_abs}), total transmittance (\ref{fig:spectrophotometer_dT}), diffuse reflectance (\ref{fig:spectrophotometer_dR}), and total reflectance (\ref{fig:spectrophotometer_tR}) where an 8$^o$ wedge is used to ensure the specular reflectance is also included.}
    \label{fig:spectrophotometer}
\end{figure}

\paragraph{Spectral measurements}\label{sec:methodsphantommeasure1}
Measurements of these phantoms are taken using a PerkinElmer Lambda 750s spectrophotometer. This dual-beam integrating sphere spectrophotometer allows for precise and accurate measurements of diffuse reflectance, total reflectance, total transmittance, and absorbance. Absorbance for each dye solution and pure dye gelatin phantom is measured with either PBS or pure gelatin as a reference respectively. The pure gelatin absorbance is also measured. 
All 1cm phantoms are used for diffuse reflectance measurements. A subset of these are also made into 5mm phantoms alongside pure intralipid gelatin 5mm phantoms, which are measured for total reflectance and total transmittance. 
All configurations are depicted for our system in Figure \ref{fig:spectrophotometer}. 

\paragraph{Inverse adding doubling (IAD)}\label{sec:methodsphantommeasure2}
IAD is used to obtain $\mu_a(\lambda)$ and $\mu_s'(\lambda)$ from total reflectance and total transmittance measurements \cite{Prahld}. We use this with the dual beam spectrophotometer setting and an incidence angle of 8$^o$ as per the experimental set-up in Figure~\ref{fig:spectrophotometer_tR}, with $g$ fixed to 0.8 for all calculations since it was found not to change the results. Since the outputted optical properties can contain a significant number of wavelengths without convergence, a Mie scattering curve is fitted to the output $\mu_s'(\lambda)$ and a second stage of IAD is run with the $\mu_s'(\lambda)$ fixed to this spectrum.
This returns outputs which are very similar but removing noise or errors.
Further details on this two-stage IAD fitting approach can be found in previous work\cite{Xie2021}. 

In order to obtain accurate optical properties from IAD a highly accurate sample depth must be provided.
To measure these accurately for the highly compressible gelatin phantoms, a CT scan is taken of each phantom using a Small Animal Radiotherapy System (SmART+, precision X-ray). A mean of 10 digital measurements is used for each phantom and an example of these measurements is shown in Figure \ref{fig:DICOM_Eg}.

Finally, a commercial calibrated phantom (BioPixS) with validated ground truth optical properties at three wavelengths is used to confirm that our output $\mu_a(\lambda)$ and $\mu_s'(\lambda)$ are correct. We highlight that only the single-stage IAD algorithm can be used here as a Mie scattering curve does not accurately represent the scattering of this sample. The BioPixS phantom is measured on three days to demonstrate the accuracy and reproducibility of IAD-outputted optical parameters.

\paragraph{Modelling tissue phantom optical properties}\label{sec:methodphantommodel}
The $\mu_a(\lambda)$ of the two-dye configuration phantoms can be modelled as in Equation \eqref{eq:phantommua} where the background spectrum is determined by measuring the absorption of pure gelatin solution, whereas the scattering is modelled based on the intralipid concentration according to trends determined using IAD. 
\begin{equation}
    \mu_{a}(\lambda) = 8\times10^{-6}c_{tot}\ln(10)[AR1 \epsilon_{AR1, eff}(\lambda) + (1 - AR14)\epsilon_{AR14, eff}(\lambda)] + \mu_{a, background}(\lambda)
    \label{eq:phantommua}
\end{equation}
In this equation $c_{tot}$ is the total concentration of dye independent of dye identity, $AR1$ and $AR14$ are the relative ratios of each dye, $\epsilon_{AR1, eff}(\lambda)$ and $\epsilon_{AR14, eff}(\lambda)$ are the effective extinction coefficients of each dye, and  $\mu_{a, background}(\lambda)$ is the measured background absorbance from gelatin. This can be adapted to a three-dye configuration as follows: 
\begin{equation}
    \mu_{a}(\lambda) = 8\times10^{-6}c_{tot}\ln(10)[AR1 \epsilon_{AR1, eff}(\lambda) + AR14\epsilon_{AR14, eff}(\lambda) + (1 - AR1 - AR14)\epsilon_{CV, eff}(\lambda)] + \mu_{a, background}(\lambda)
    \label{eq:phantommua3}
\end{equation}
Where $CV$ is the relative ratio of crystal violet and $\epsilon_{CV, eff}(\lambda)$ is the effective extinction coefficient of this dye.

\subsection{Evaluation of model performance}\label{sec:methodevaluate}
In this work, the above models are evaluated against spectra with known ground truth. These reference spectra are either found by Monte Carlo simulation or measurement of controlled phantoms, however the evaluation of each is broadly the same. The forward models are evaluated in terms of the fit of their predicted spectra, and the inverse problem solutions (parametric fits) are investigated in terms of the quality of parameter extraction. 

Diffuse reflectance spectra can be predicted by inputting ground truth values into each forward tissue model and compared to corresponding reference spectra. The Normalised Root Mean Squared Error ($NRMSE$), defined in Equation \eqref{eq:NRMSE}, is calculated to quantitatively evaluate the similarity between the reflectance predicted by the forward model and the associated reference reflectances. In the context of this work the wavelength ranges considered for this metric correspond to those ranges used for fitting the inverse models.
\begin{equation}
    NRMSE = \frac{\sqrt{\frac{1}{\Lambda}\sum_{\lambda}^{\Lambda}\left(s_{\lambda} - r_{\lambda}\right)^2}}{\sqrt{\frac{1}{\Lambda}\sum_{\lambda}^{\Lambda} r^2_{\lambda}}}
    \label{eq:NRMSE}
\end{equation}
Here the modelled spectrum $s_{\lambda}$ at any given wavelength $\lambda$ is evaluated against the intensity at the same wavelength in the reference spectrum $r_{\lambda}$ using the normalized root mean squared error ($NRMSE$) calculated across all $\Lambda$ wavelengths. These forward model reflectance spectra are also plotted against the reference reflectance spectra and a regression line calculated. This is evaluated using the Pearson correlation coefficient ($r$) and the p-value ($p$) for a hypothesis test with the null hypothesis that there is no correlation. Using a 95\% confidence interval, significant $p$ is less than 0.05, whereas a strong correlation is demonstrated by $r$ close to 1.  This gives an indication of similarity in shape, while disregarding offsets.

The inverse problem solutions can also be evaluated by determining how well they recover the ground truth input parameters using a non-linear least-squares fitting approach (using SciPy v1.10.0 \href{https://docs.scipy.org/doc/scipy/reference/generated/scipy.optimize.least_squares.html}{\texttt{scipy.optimize.least\_squares}} function). The cost function for this is shown in Equation \eqref{eq:leastsquares} where $R(\lambda)$ is the reference spectrum and $M$ is the forward model function. This equation is modified to fit the parameters used to calculate $\mu_a(\lambda)$ and $\mu_s'(\lambda)$ directly.
\begin{equation}
    argmin_{\mu_{a}(\lambda), \mu_{s}'(\lambda)} \sum (R - M(\mu_{a}(\lambda), \mu_s'(\lambda))^2
    \label{eq:leastsquares}
\end{equation}

All models are fitted by only considering wavelengths up to 600nm as this is the region the Yudovsky model has been shown to be effective by the authors\cite{Yudovsky2011a}.
When fitting to measured phantom spectra, only wavelengths up to 575nm are considered due to the forward model predicted spectra being highly unreliable after this point for all three models as discussed in Section \ref{sec:resultsPhantoms}. The quality of parameter recovery is examined by calculating the correlation between the fitted and ground truth parameters (using SciPy v1.10.0 \href{https://docs.scipy.org/doc/scipy/reference/generated/scipy.stats.linregress.html}{\texttt{scipy.stats.linregress}} function). This is evaluated with the Pearson correlation coefficient ($r$) and the p-value ($p$) as above. The quality of parameter recovery is also evaluated using absolute percentage errors ($APE$) as given by Equation \eqref{eq:percenterr} where $e$ is the extracted parameter and $g$ is the ground truth parameter. The median and inter-quartile range of these parameters are presented for each dataset. 
\begin{equation}
    APE = |\frac{e - g}{g}| \times 100 
    \label{eq:percenterr}
\end{equation}

These evaluation methods can be done for quantitative or relative spectra. Here relative spectra are defined as mean normalised spectra. This allows parameters to be extracted considering the shape of the spectrum but without considering absolute intensity. This is investigated as it is simpler to capture relative data in clinical environments \cite{Bahl2023}.

\section{Results}\label{sec:results}
\subsection{Monte Carlo}\label{sec:resultsMC}
Each model has hyperparameters which are fitted to Monte Carlo datasets for refractive indices 1.33, 1.35, and 1.44. These are listed in Appendix Table \ref{tb:fittedmodelparams} alongside any literature hyperparameters \cite{Jacques1999, Yudovsky2009}. It should be noted that the literature hyperparameters for Jacques could not be
directly
replicated by fitting to our Monte Carlo simulations. 
Refitting these hyperparameters leads to the mean ($\pm$ standard deviation) $NRMSE$ of the $n=1.33$ dataset improving from 0.080($\pm$0.056) to 0.021($\pm$0.028). 
In contrast, Yudovsky's literature hyperparameters are similar to our refitting.
For Yudovsky, despite the expected equivalence, we observed a small discrepancy in fitting quality between the ``extensive model'' \cite{Yudovsky2009} and the ``simplified model'' described in their Erratum\cite{Yudovsky2015}.
This can be seen for a refractive index of 1.44 in Appendix Figure \ref{fig:badYudovsky}, which matches the results quoted in their Erratum\cite{Yudovsky2015}.
The ``simplified model'' improves the mean ($\pm$ standard deviation) $NRMSE$ of the $n=1.44$ dataset from 0.050($\pm$0.014) to 0.010($\pm$0.003).
For this reason the Erratum model is used for this work. 

\begin{table}[bhp]
    \centering
    \caption{Mean ($\pm$ standard deviation) $NRMSE$ (3.d.p.) between each forwards spectrum from each model and each of 100 Monte Carlo simulated spectra using the same ground truth variable parameters for each refractive index dataset and each analytical model. This is presented with the Pearson $r$ (bold if Pearson $p < 0.05$) for the linear regression between all forwards spectra against Monte Carlo simulated spectra for each refractive index dataset and each analytical model. All metrics are evaluated for the wavelength region of 450-600nm.}
    \begin{tabular}{|c|c|c|c|}
        \hline
        Model & Refractive index & $NRMSE$ & $r$ \\
        \hline
        \multirow{3}{*}{Yudovsky 2009} & 1.33 & 0.013 ($\pm$ 0.006) & \textbf{1.000} \\
        & 1.35 & 0.013 ($\pm$ 0.005) & \textbf{1.000} \\
        & 1.44 & 0.010 ($\pm$ 0.004) & \textbf{1.000} \\
        \hline
        \multirow{3}{*}{Jacques 1999} & 1.33 & 0.037 ($\pm$ 0.065) & \textbf{0.999} \\
        & 1.35 & 0.045 ($\pm$ 0.074) & \textbf{0.999} \\
        & 1.44 & 0.030 ($\pm$ 0.048) & \textbf{1.000} \\
        \hline
        \multirow{3}{*}{Modified Beer-Lambert} & 1.33 & 0.630 ($\pm$ 0.461) & \textbf{0.667} \\
        & 1.35 & 0.603 ($\pm$ 0.339) & \textbf{0.639} \\
        & 1.44 & 0.675 ($\pm$ 0.620) & \textbf{0.605} \\
        \hline
    \end{tabular}
    \label{tb:NRMSEsingle}
\end{table}

\begin{figure}[htbp]
    \centering
    \begin{subfigure}{0.49\textwidth}
        \includegraphics[width=\textwidth]{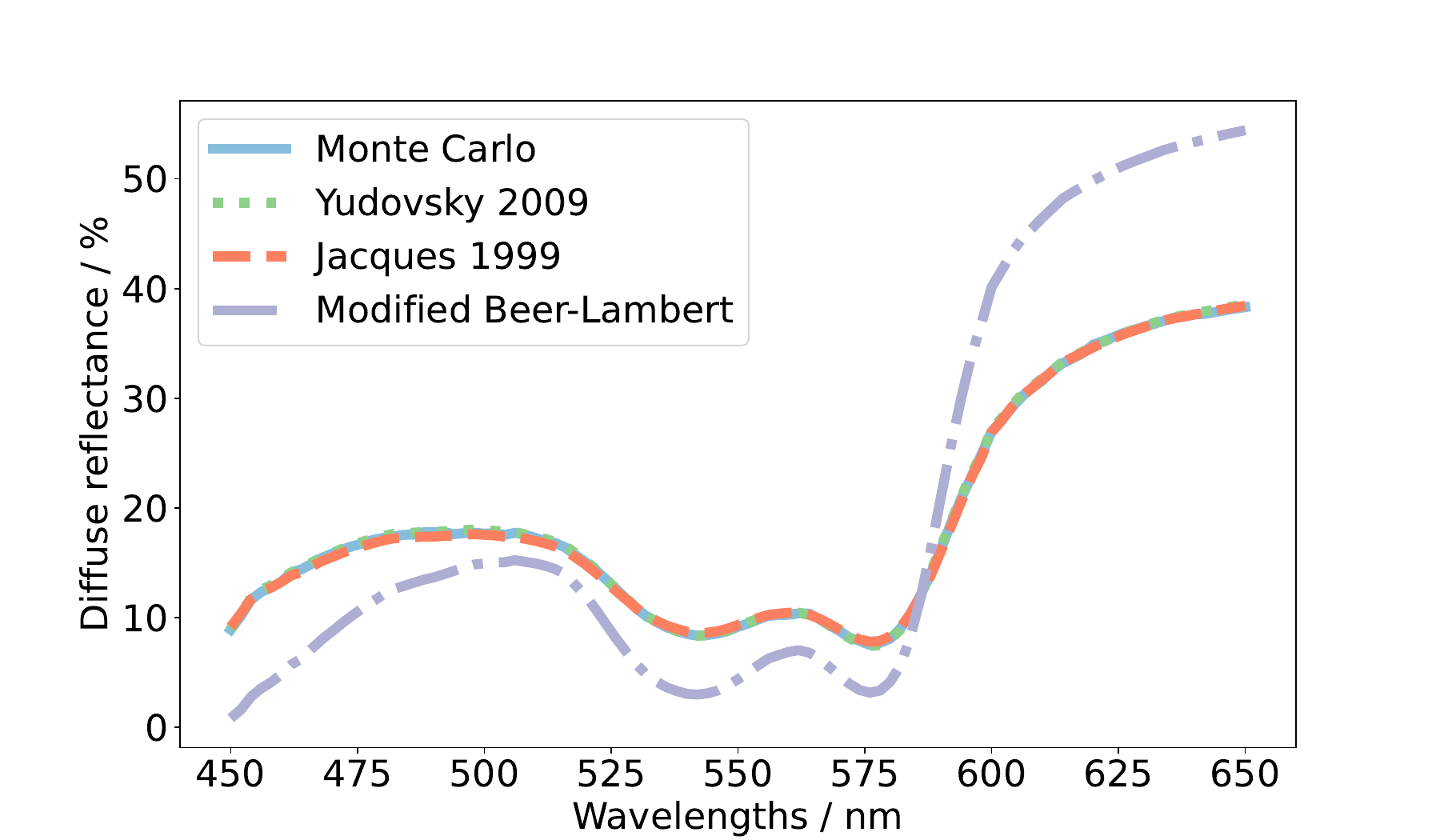}
        \caption{}
        \label{fig:egspectrasingle}
    \end{subfigure}
    \begin{subfigure}{0.49\textwidth}
        \includegraphics[width=\textwidth]{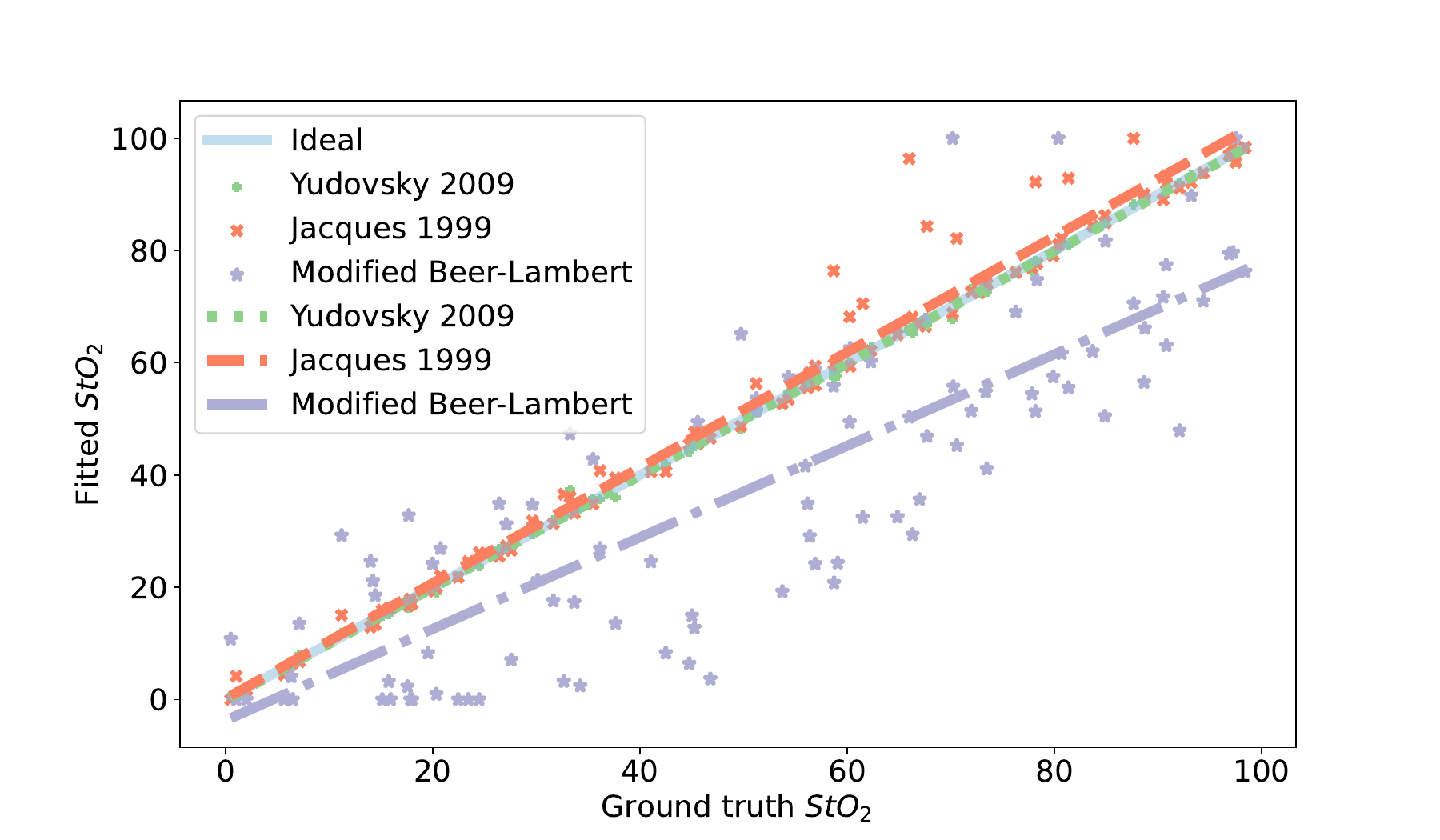}
        \caption{}
        \label{fig:egtrendsingle}
    \end{subfigure}
    \caption{Figure depicting evaluation of forward models (\ref{fig:egspectrasingle}) and inverse problem solutions (\ref{fig:egtrendsingle}). 
    Figure \ref{fig:egspectrasingle} depicts an example of the predicted spectra from each forward analytical model: Yudovsky 2009 (\textcolor{MyGreen}{green dotted}), Jacques 1999 (\textcolor{MyOrange}{orange dashed}), and Modified Beer-Lambert (\textcolor{MyPurple}{purple dot-dashed}), using ground truth variables for a refractive index of 1.44 compared to that predicted by Monte Carlo (\textcolor{MyBlue}{blue solid}).
    Figure \ref{fig:egtrendsingle} shows an example of difference in quality of parameter recovery by fitting each inverse analytical model to Monte Carlo simulations in the wavelength range of 450-600nm: Yudovsky 2009 (\textcolor{MyGreen}{green $+$}), Jacques 1999 (\textcolor{MyOrange}{orange $\times$}), and Modified Beer-Lambert (\textcolor{MyPurple}{purple $*$}), and their associated trend lines for a refractive index of 1.44 for $StO_2$.}
    \label{fig:FwDandInverse}
\end{figure}

Example data comparing the analytical models to Monte Carlo simulations can be seen in Figure \ref{fig:FwDandInverse}.
For each model, spectra are generated with each
input
parameter set from the Monte Carlo dataset and $NRMSE$ is calculated per spectrum. The mean ($\pm$ standard deviation) of these $NRMSE$ per model per refractive index are shown in Table~\ref{tb:NRMSEsingle}.
Only wavelengths until 600nm are considered for these metrics as in Section \ref{sec:methodevaluate}. An example spectrum with each of the forward models and modelled spectra using ground truth parameters are shown in Figure \ref{fig:egspectrasingle} for a refractive index of 1.44. A regression line calculated between each forward model dataset and the Monte Carlo simulated dataset for each refractive index and the correlation coefficient ($r$) can be seen in Table \ref{tb:NRMSEsingle}.

\begin{table}[htpb]
    \centering
    \caption{The  Pearson $r$ (bold if $p<0.05$) of the linear regression line between the fitted tissue parameters and their ground truth displayed with their median (inter-quartile range) absolute percentage errors ($APE$). This is shown for each variable and for a refractive index of 1.44 when extracted by fitting Yudovsky 2009 (Y), Jacques 1999 (J), or Modified Beer-Lambert (BL) to the Monte-Carlo dataset in the wavelength range 450-600nm. All presented to 3s.f.}
    \begin{tabular}{|cc|cc|}
        \hline
        parameter & model & $r$ & median (inter-quartile range) \\
        & &  & $APE$ (\%)\\
        \hline
        \multirow{3}{*}{$StO_2$} & Y & \textbf{1.00} & 0.913 (1.92) \\
        & J & \textbf{0.986} & 2.21 (4.97) \\
        & BL & \textbf{0.838} & 43.0 (49.8) \\
        \hline
        \multirow{3}{*}{$f_{blood}$} & Y & \textbf{0.982} & 5.68 (6.08) \\
        & J & \textbf{0.928} & 7.26 (16.2) \\
        & BL & \textbf{0.582} & 49.6 (32.2) \\
        \hline
        \multirow{3}{*}{$a$} & Y & \textbf{0.992} & 3.90 (4.73) \\
        & J & \textbf{0.959} & 4.54 (15.3) \\
        & BL & \textbf{-0.706} & 40.0 (114) \\
        \hline
        \multirow{3}{*}{$b$} & Y & \textbf{1.00} & 1.50 (2.92) \\
        & J & \textbf{0.963} & 2.86 (9.11) \\
        & BL & \textbf{-0.446} & 95.4 (5.68) \\
        \hline
    \end{tabular}
    \label{tb:singleparamtrends}
\end{table}

Finally, the inverse problems with our fixed hyperparameters are fitted as in \ref{sec:methodevaluate} to this same Monte Carlo dataset to recover the $StO_2$, $f_{blood}$, $a$, and $b$ tissue parameters. A linear regression is fitted between these retrieved values and the ground truth counterparts. The Pearson $r$ values of this, and the median (inter-quartile range) absolute percentage errors between the fitted and ground truth tissue parameters are shown in Table \ref{tb:singleparamtrends} for a refractive index of 1.44 with further parameters and refractive indices shown in Appendix Table \ref{tb:singleparamtrendsfull}. An example of the fitted parameters compared to the ground truth parameters is shown for $StO_2$ at a refractive index of 1.44 in Figure \ref{fig:egtrendsingle}.


\subsection{Gelatin-based tissue phantoms}\label{sec:resultsPhantoms}
\begin{figure}[htbp]
    \centering
    \begin{subfigure}{0.49\textwidth}
        \includegraphics[width=\textwidth]{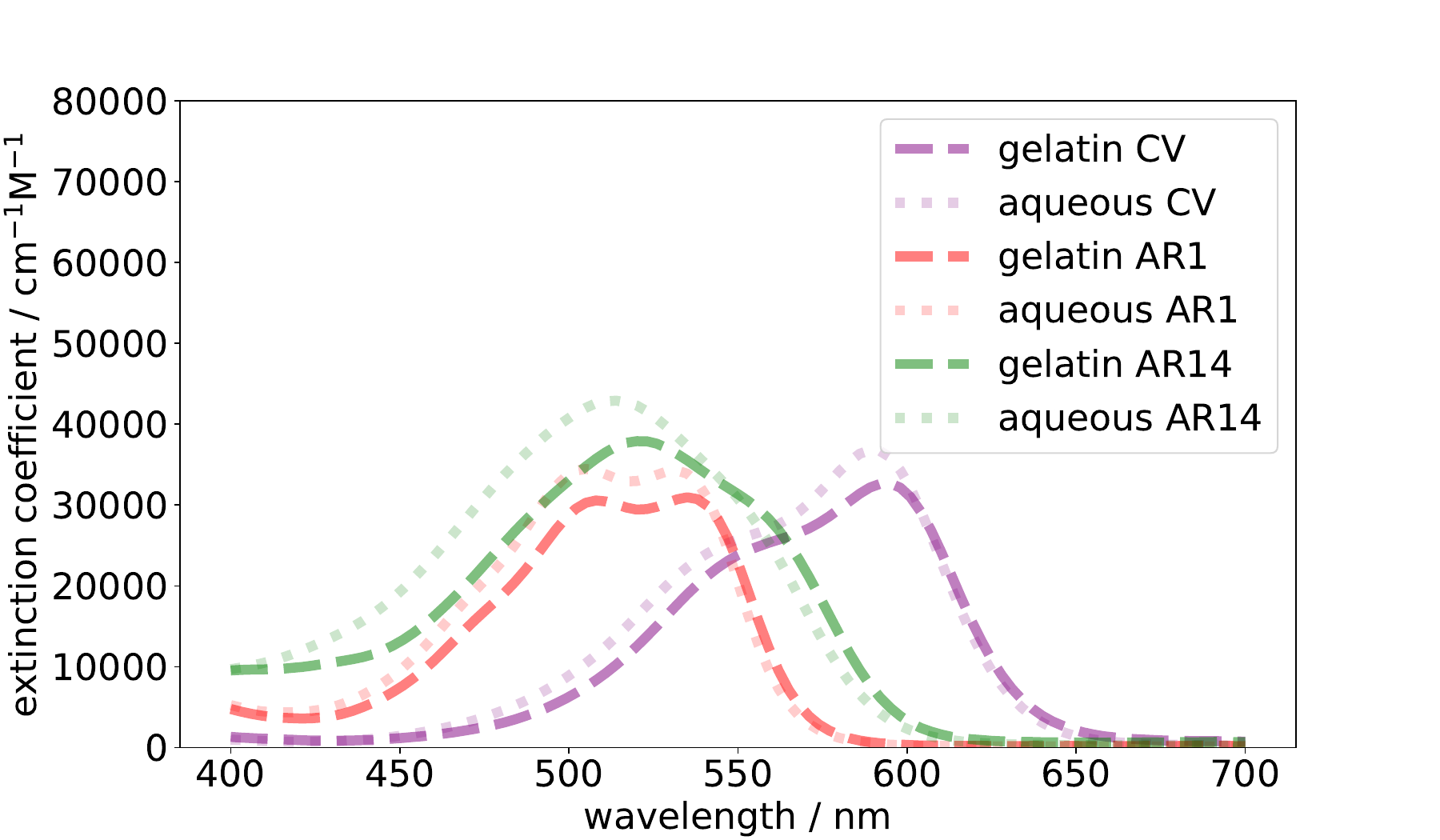}
        \caption{}
        \label{fig:epseff}
    \end{subfigure}
    \begin{subfigure}{0.49\textwidth}
        \includegraphics[width=\textwidth]{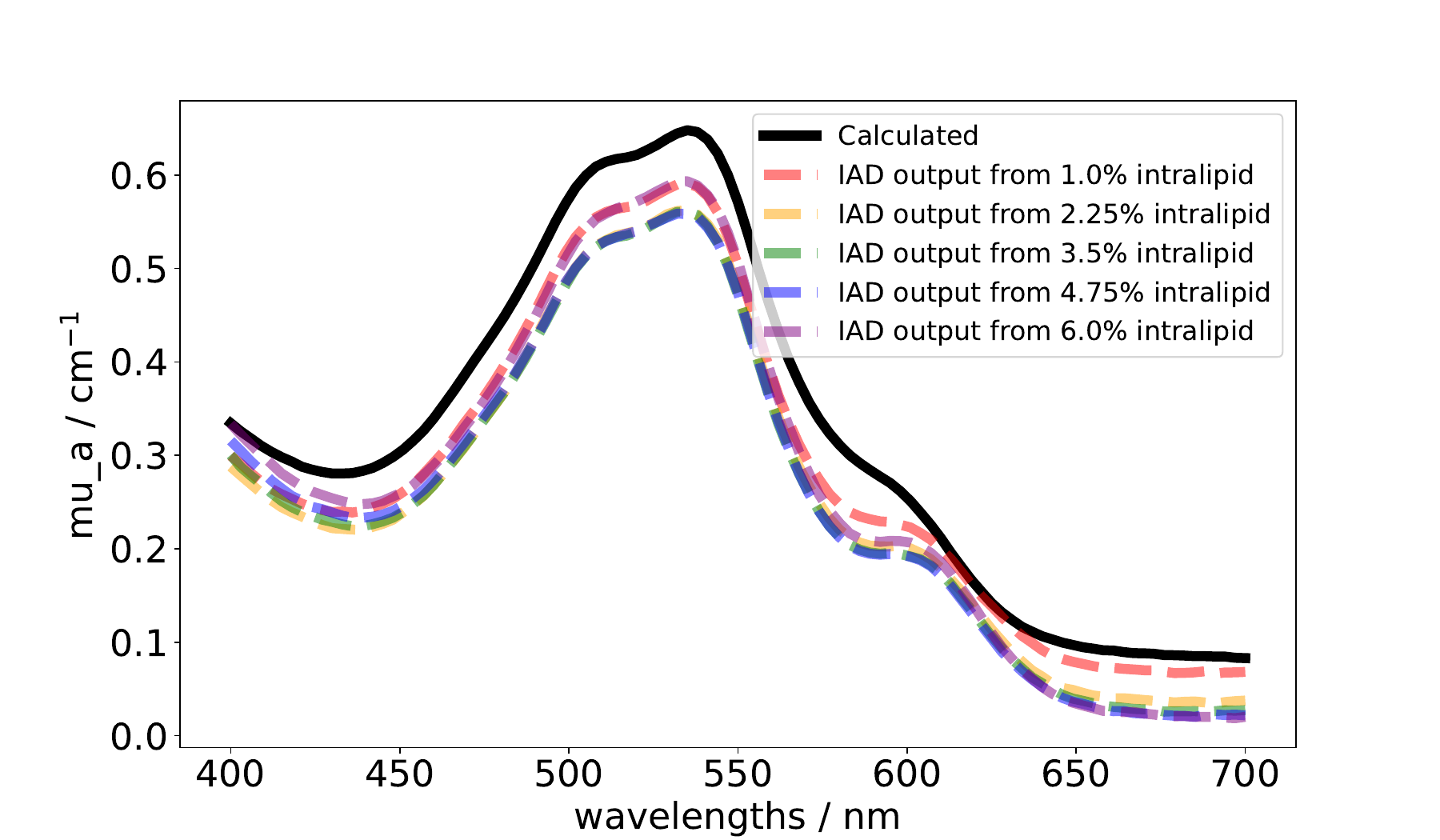}
        \caption{}
        \label{fig:IADmua}
    \end{subfigure}
    \caption{Figure \ref{fig:epseff} shows $\epsilon_{eff}(\lambda)$ calculated for each dye [acid red 1 (AR1), acid red 14 (AR14), crystal violet (CV)] measured in gelatin (dashed) or PBS (dotted) demonstrating the shift in peaks. Figure \ref{fig:IADmua} shows a calculated (black solid) $\mu_a(\lambda)$ for one 3-dye configuration compared to the IAD output $\mu_a(\lambda)$ (coloured dashed) from measurements of phantoms in this configuration with a range of intralipid concentrations.}
    \label{fig:Phantombackgrounds}
\end{figure}

Each dye absorbance is measured in both an aqueous and gelatin based solution with effective extinction coefficients calculated in each case. Whilst the aqueous measurements closely match the expected $\epsilon_{eff}(\lambda)$ from literature\cite{PhotochemCAD}, the gelatin measurements show a shifting of peaks, as seen in Figure~\ref{fig:epseff}. This is likely due to interaction of the dyes with gelatin altering their interactions with solvent as has been noted with other dyes \cite{Cook2011}.
Since these peak shifts are seen in all subsequent data, these gelatin-based $\epsilon_{eff}(\lambda)$ are used in the model analysis. A pure gelatin solution absorption is measured at the concentration used in each phantom. This is used as a background $\mu_a(\lambda)$, as seen in Appendix Figure~\ref{fig:muaback}, to account for any absorption not due to dyes.
We further assume that intralipid is a purely scattering medium. This is considered a reasonable assumption since the IAD returned $\mu_a$ for the purely intralipid phantoms are similar to the pure gelatin background $\mu_a$ seen in Appendix Figure~\ref{fig:muaback}.

IAD\cite{Prahld} is used to analyse phantoms of 5mm depth with or without dyes in all intralipid concentrations. The $a$ and $b$ Mie coefficients fitted to the scattering in these cases are plotted against intralipid concentration. It was found that $b$ had no significant trend with intralipid concentration and so a median value of 0.98 was used in this work.
However, a clear trend was identified for $a$ as can be seen in Appendix Figure \ref{fig:atrend}. 
This trend is described in Equation \eqref{eq:atrend}, where $I$ is the intralipid concentration and $I_0$ is the reference concentration of 1\%v/v: 
\begin{equation}
    a = 6.66cm^{-1}\frac{I}{I_0} + 2.55cm^{-1}
    \label{eq:atrend}
\end{equation}
A Pearson correlation coefficient ($r$) of 0.99 and p of 0.00 was observed, therefore this is used throughout this work to model scattering based on Intralipid concentration.

The calculated $\mu_a(\lambda)$ using $\epsilon_{eff}(\lambda)$ displayed in Figure \ref{fig:epseff} is compared to the IAD outputs of phantoms including dyes with each intralipid concentration.
An example of this is seen in Figure \ref{fig:IADmua}. This shows the overall spectrum is accurate, however the peak for CV (approx 600nm) appears shifted in wavelength, likely due to interactions with intralipid which cannot be captured in absorbance measurements. 

\begin{figure}[htbp]
    \centering
    \begin{subfigure}{0.33\textwidth}
        \includegraphics[width=\textwidth]{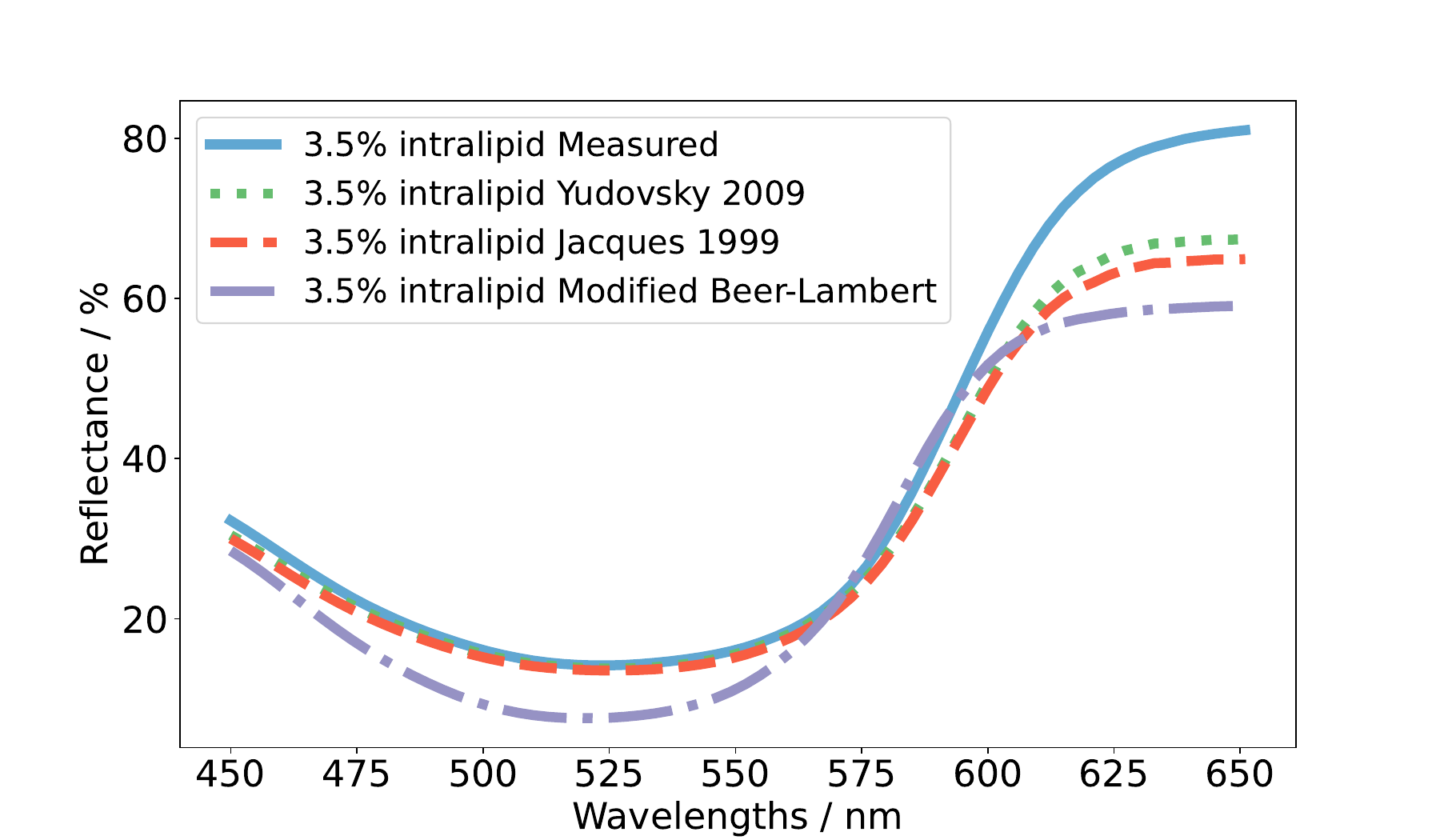}
        \caption{}
        \label{fig:phantomforwardsquant}
    \end{subfigure}
    \begin{subfigure}{0.33\textwidth}
        \includegraphics[width=\textwidth]{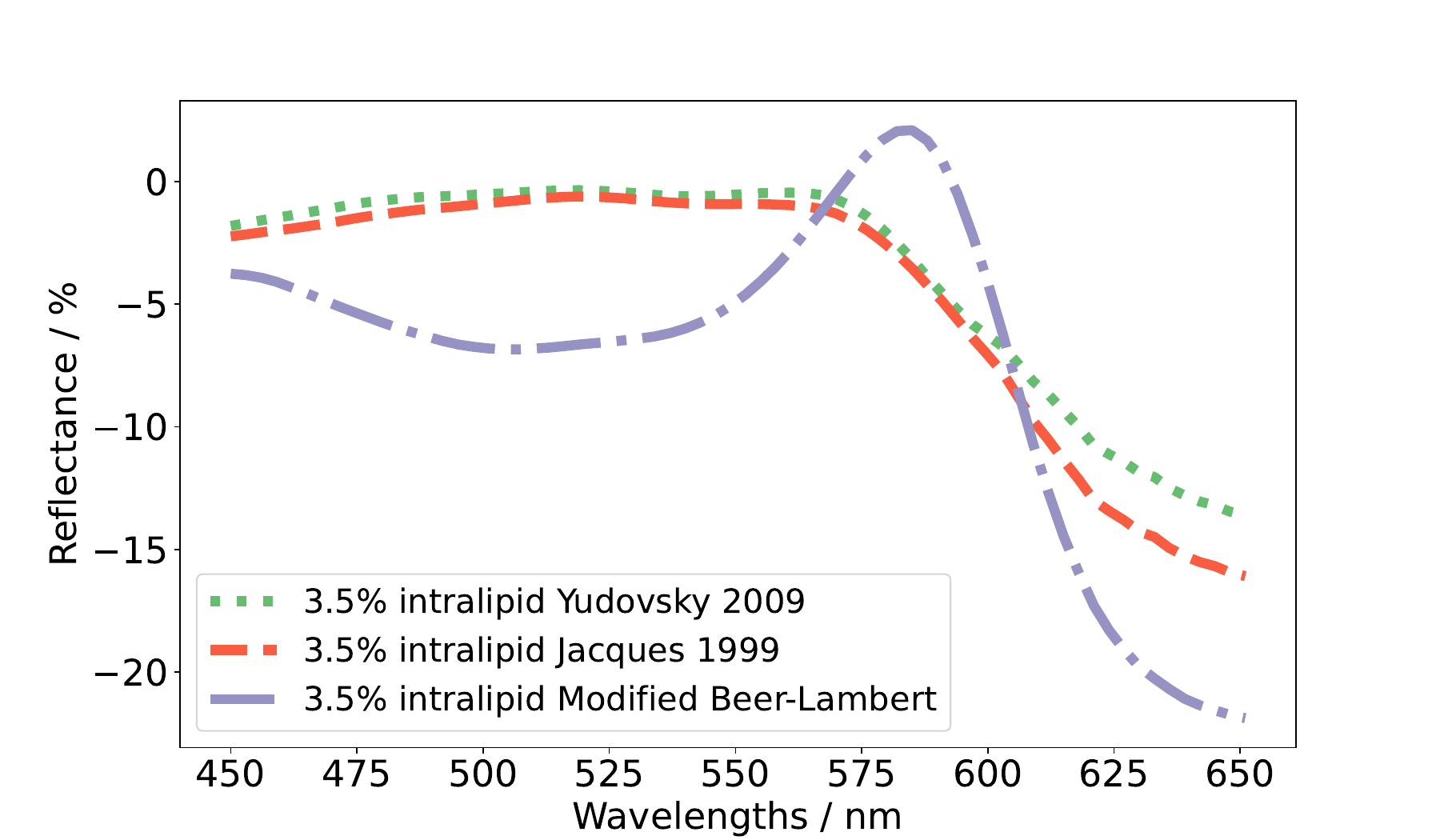}
        \caption{}
        \label{fig:phantomresidual}
    \end{subfigure}
    \begin{subfigure}{0.33\textwidth}
        \includegraphics[width=\textwidth]{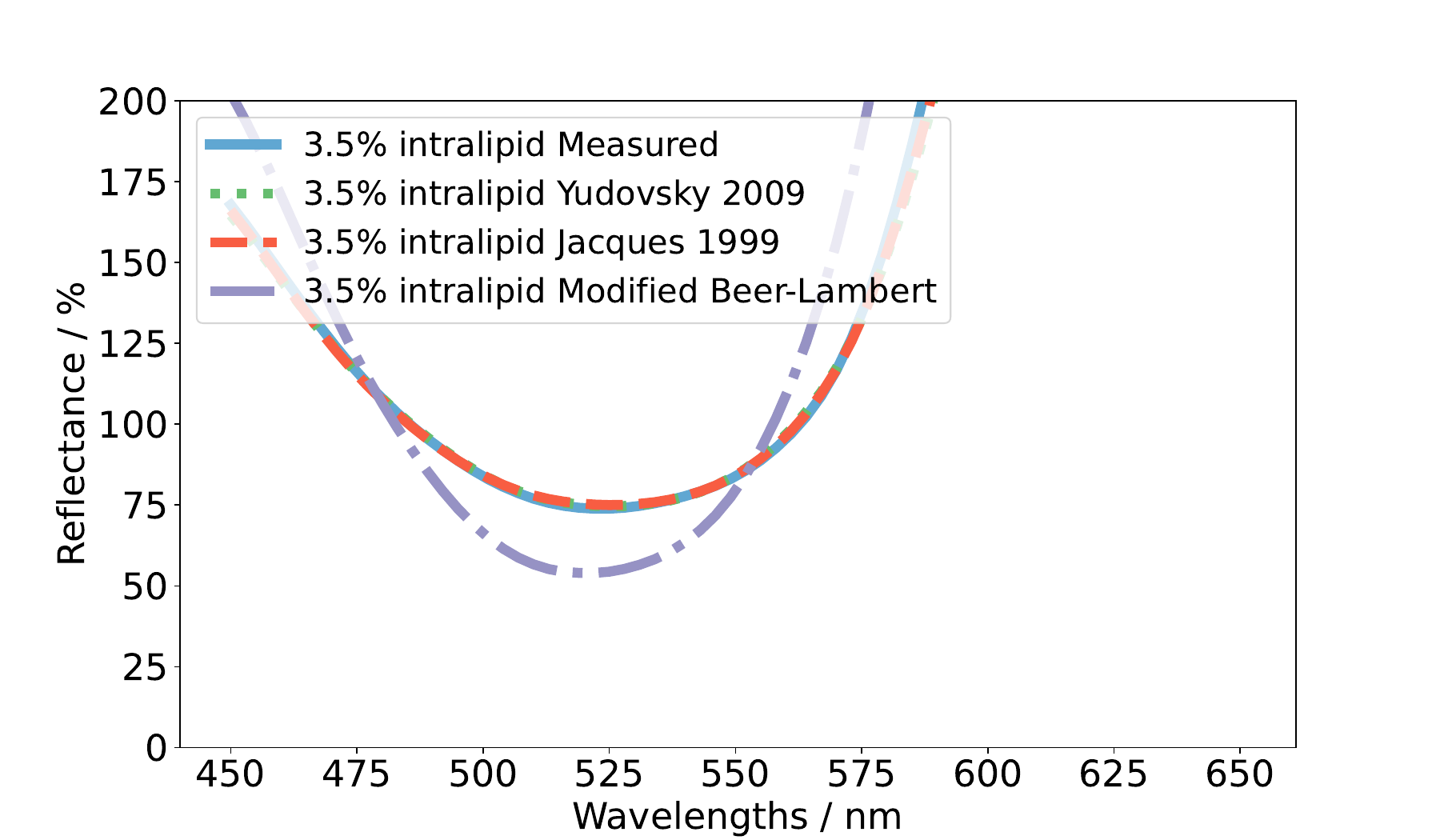}
        \caption{}
        \label{fig:phantomforwardsnorm}
    \end{subfigure}
    \hfill
    \begin{subfigure}{0.49\textwidth}
        \includegraphics[width=\textwidth]{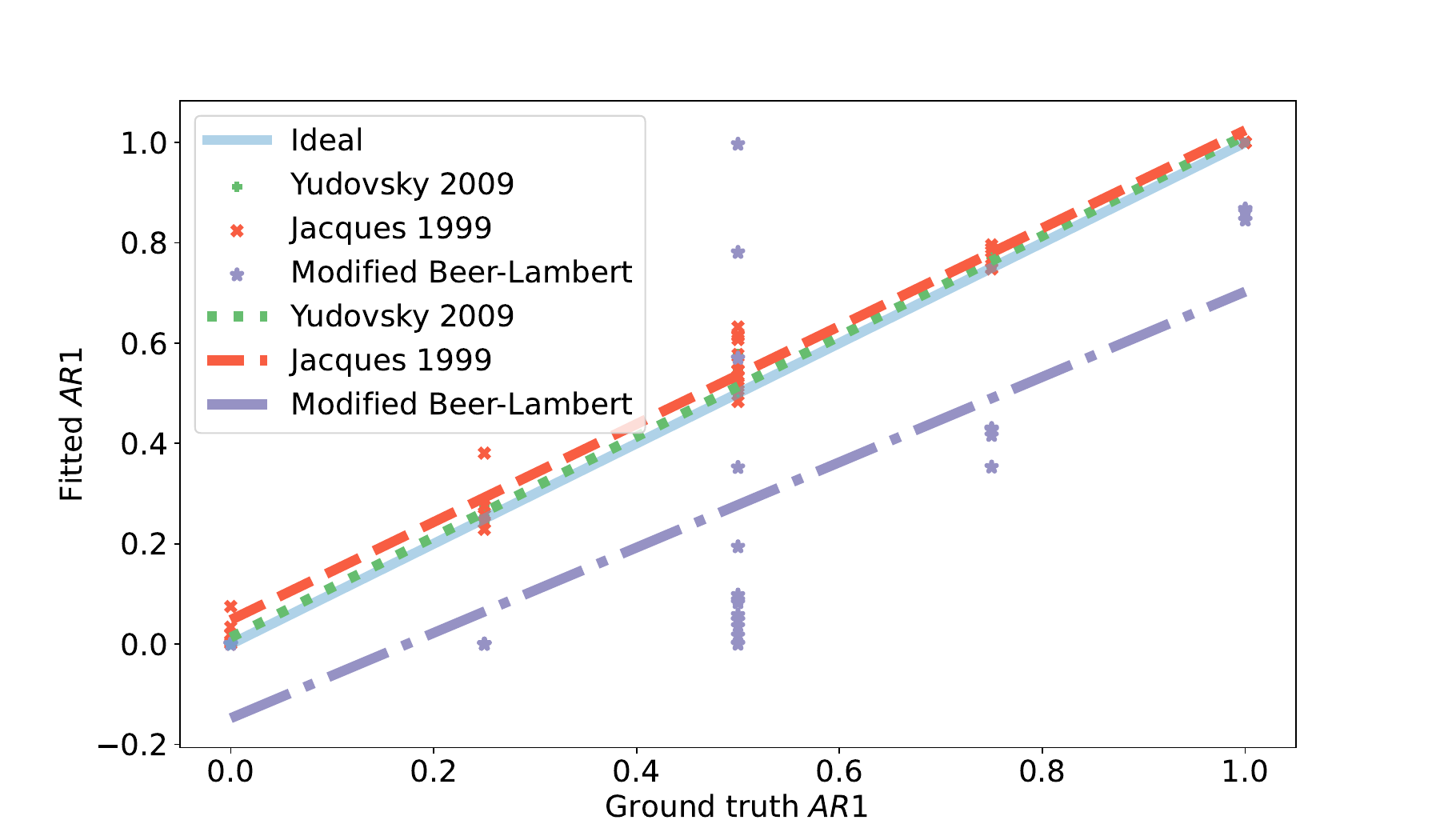}
        \caption{}
        \label{fig:phantomAR1quant}
    \end{subfigure}
    \begin{subfigure}{0.49\textwidth}
        \includegraphics[width=\textwidth]{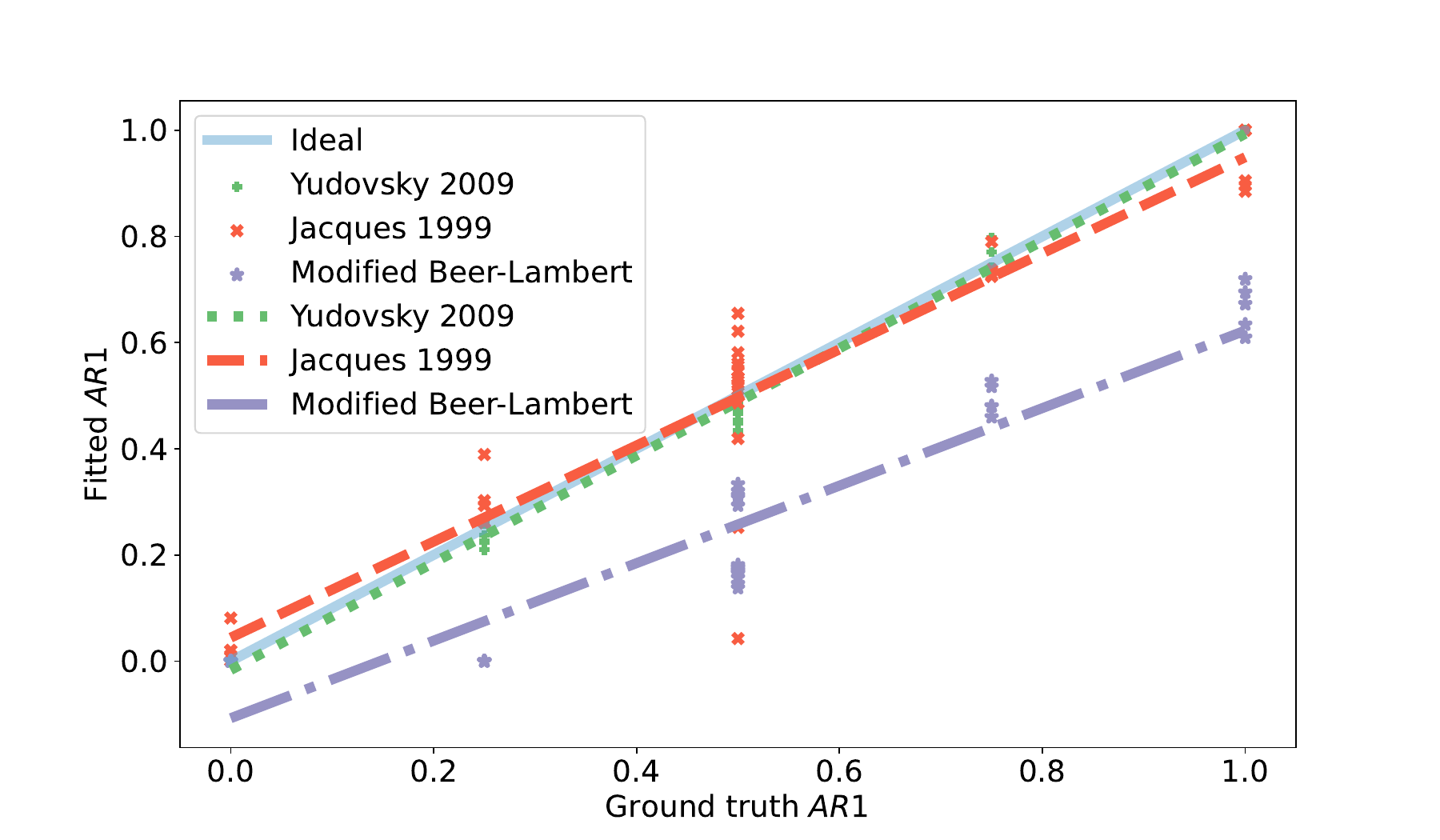}
        \caption{}
        \label{fig:phantomAR1norm}
    \end{subfigure}
    \caption{Example of measured (\textcolor{MyBlue}{blue}) spectrum for one 2-dye configuration at an intralipid concentration of 3.5\% compared to the respective spectra generated from the Yudovsky 2009 (\textcolor{MyGreen}{green dotted}), Jacques 1999 (\textcolor{MyOrange}{orange dashed}), and Modified Beer-Lambert (\textcolor{MyPurple}{purple dot-dashed}) models using the ground truth parameters for quantitative (\ref{fig:phantomforwardsquant}) or relative (\ref{fig:phantomforwardsnorm}) spectra (mean normalised in the range 450-575nm). The residuals between the modelled quantitative spectra and the measured spectra are shown in \ref{fig:phantomresidual}. Examples of the parameter recovery from the measured phantom spectra for the AR1 proportion by Yudovsky 2009 (\textcolor{MyGreen}{green +}), Jacques 1999 (\textcolor{MyOrange}{orange $\times$)}, and Modified Beer-Lambert (\textcolor{MyPurple}{purple $*$}) models compared with the ground truth parameters (\textcolor{MyBlue}{blue solid}) for fits to quantitative (\ref{fig:phantomAR1quant}) or relative (\ref{fig:phantomAR1norm}) spectra in the wavelength range 450-575nm.}
 \label{fig:phantomforwards}
\end{figure}

Using a refractive index of 1.35\cite{Pogue2006} diffuse reflectance spectra are generated from each model using $\mu_a(\lambda)$ and $\mu_s'(\lambda)$ from Equations~\eqref{eq:phantommua} and \eqref{eq:Mie} using the trend in Equation \eqref{eq:atrend} and the median $b$ value for each 2-dye configuration and intralipid concentration. The forward Yudovsky, Jacques, and Modified Beer-Lambert models in Section \ref{sec:resultsMC} are compared to diffuse reflectance measurements for each phantom.
An example of this can be seen in Figure \ref{fig:phantomforwardsquant} for quantitative spectra or Figure \ref{fig:phantomforwardsnorm} for mean normalised relative data.
The data is normalised using the mean of the wavelength range 450-575nm as beyond this region the models appear to fit poorly as seen in Figure \ref{fig:phantomresidual}. 
The mean $NRMSE$ ($\pm$ standard deviation) between each spectrum and the measured spectrum, in the region 450-575nm, for this dataset can be seen in Table \ref{tb:NRMSEphantom}. 
The Pearson correlation coefficient $r$ between the forwards spectra compared to the measured spectra for each model is shown in Table \ref{tb:NRMSEphantom} and further parameters in Appendix Table \ref{tb:phantomparamsfull}. Whilst the overall magnitude of these is higher, the trend in quality of fit is similar as to the Monte Carlo simulations. 

The inverse problems are solved by non-linear least squares  approaches using measured spectrum as reference observations. The recovered parameters are correlated to the ground truth parameters. The associated errors and correlation coefficients can be seen in Table \ref{tb:NRMSEphantom} (and further in Appendix Table \ref{tb:phantomparamsfull}) and an example of the fitted parameters compared to ground truth parameters is shown for $AR1$ in Figure~\ref{fig:phantomAR1quant} for fits to quantitative spectra or Figure~\ref{fig:phantomAR1norm} for fits to 
relative spectra.

\begin{table}[htbp]
    \centering
    \caption{mean ($\pm$ standard deviation) $NRMSE$ (3.d.p.) between the modelled spectrum using the ground truth parameters and the measured spectrum of a tissue phantom for each analytical model for either quantitative or relative spectra. Alongside this is the Pearson $r$ (bold if $p < 0.05$) for each model for all forwards spectra compared to the measured spectra. All metrics evaluated for the wavelength range 450-575nm}
    \begin{tabular}{|c|c|c|c|}
        \hline 
        Model & Quantitative (Q) or Relative (R) & $NRMSE$ & $r$ \\
        \hline
        \multirow{2}{*}{Yudovsky 2009} & Q & 0.059 ($\pm$ 0.031) & \textbf{0.998} \\
        & R & 0.012 ($\pm$ 0.009) & \textbf{0.999} \\
        \hline
        \multirow{2}{*}{Jacques 1999} & Q & 0.075 ($\pm$ 0.032) & \textbf{0.998} \\
        & R & 0.022 ($\pm$ 0.023) & \textbf{0.995} \\
        \hline
        \multirow{2}{*}{Modified Beer-Lambert} & Q & 0.464 ($\pm$ 0.294) & \textbf{0.808} \\
        & R & 0.215 ($\pm$ 0.139) & \textbf{0.930} \\
        \hline
    \end{tabular}
    \label{tb:NRMSEphantom}
\end{table}


\begin{table}[htbp]
    \centering
    \caption{The  Pearson $r$ (bold if $p<0.05$) of the linear regression line between the fitted parameters and their ground
    truth displayed with their median (inter-quartile range) absolute percentage errors for each variable when extracted by fitting Yudovsky 2009 (Y), Jacques 1999 (J), or Modified Beer-Lambert (BL) for both quantitative and relative data. All metrics calculated for the wavelength range 450-575nm and presented to 3s.f.}
    \begin{tabular}{|ccc|cc|}
        \hline
        Parameter & Model & Quantitative (Q) & $r$ & median (inter-quartile range) \\
        & & or Relative (R) & & $APE$ (\%)\\
        \hline
        \multirow{6}{*}{$AR1$} & \multirow{2}{*}{Y} & Q & \textbf{0.997} & 1.59 (11.0) \\
        & & R & \textbf{0.998} & 4.42 (8.31) \\
        \cline{2-5}
        & \multirow{2}{*}{J} & Q & \textbf{0.989} & 7.02 (20.2) \\
        & & R & \textbf{0.934} & 10.4 (23.7) \\
        \cline{2-5}
        & \multirow{2}{*}{BL} & Q & \textbf{0.756} & 83.5 (57.0) \\
        & & R & \textbf{0.939} & 50.0 (31.4)\\
        \hline
        \multirow{6}{*}{$AR14$} & \multirow{2}{*}{Y} & Q & \textbf{1.00} & 1.36 (8.69) \\
        & & R & \textbf{0.998} & 2.48 (5.76) \\
        \cline{2-5}
        & \multirow{2}{*}{J} & Q & \textbf{0.989} & 7.02 (16.3) \\
        & & R & \textbf{0.934} & 6.34 (10.0)\\
        \cline{2-5}
        & \multirow{2}{*}{BL} & Q & \textbf{0.756} & 80.4 (72.1) \\
        & & R & \textbf{0.939} & 37.5 (66.7) \\
        \hline
        \multirow{6}{*}{$I$} & \multirow{2}{*}{Y} & Q & \textbf{0.648} & 125 (276) \\
        & & R & 0.310 & 93.9 (30.0) \\
        \cline{2-5}
        & \multirow{2}{*}{J} & Q & \textbf{0.540} & 122 (288) \\
        & & R & \textbf{-0.451} & 100 (55.7) \\
        \cline{2-5}
        & \multirow{2}{*}{BL} & Q & \textbf{-0.705} & 100 (172) \\
        & & R & -0.157 & 100 (0.00) \\
        \hline
        \multirow{6}{*}{$c_{tot}$} & \multirow{2}{*}{Y} & Q &  \textbf{0.500} & 102 (226) \\
        & & R & 0.303 & 66.1 (46.8) \\
        \cline{2-5}
        & \multirow{2}{*}{J} & Q & \textbf{0.684} & 100 (260) \\
        & & R & -0.0959 & 89.7 (267)\\
        \cline{2-5}
        & \multirow{2}{*}{BL} & Q & \textbf{0.721} & 35.2 (21.6) \\
        & & R & \textbf{0.689} & 44.6 (23.6)\\
        \hline
    \end{tabular}
    \label{tb:phantomparams}
\end{table}

Finally, some 3-dye configurations are investigated with the addition of CV. Some examples of both quantitative and relative spectral reconstruction using the ground truth parameters in the forward Yudovsky model for a single 3-dye configuration at a range of intralipid concentrations can be seen in Appendix Figure \ref{fig:3phantomforwards}. Despite considering the shift in CV peak due to gelatin, there appears to be a further shift, likely due to interactions with intralipid. This can be seen as an offset between the measured and reconstructed spectra in Appendix Figure \ref{fig:3phantomforwards} and appears to change with intralipid concentration. Due to the limited ground truth parameter range for these 3-dye configurations, only median percentage errors are presented for the inverse problem solution performance in Appendix Table \ref{tb:3phantomparams} alongside their inter-quartile range. These show a dramatic loss in accuracy of $AR1$ and $AR14$ recovery, despite good $CV$ recovery, likely due to the imperfect $CV$ absorbance modelling distorting the spectrum. 

\section{Discussion and future work}\label{sec:discussion}
When compared to Monte Carlo simulations, the semi-empirical Yudovsky model performs best in terms of fit and parameter extraction, followed closely by the Jacques model. The Modified Beer-Lambert model, however, performs much more poorly. It is also noted that the Jacques model fits Monte Carlo simulations significantly better after refitting the hyperparameters, however the Yudovsky model performs well using literature values provided that the simplified erratum model is used. 

When modelling tissue phantom spectra, it is important to have some prior knowledge of the phantoms. This allows accurate $\mu_a(\lambda)$ and $\mu_s'(\lambda)$ calculation, for example utilising the shift in dye peaks from their literature data, background gelatin absorbance, and the trend in Mie scattering parameters with intralipid concentration. Some variation is seen in IAD which may be due to inaccuracies in sample depth measurement or variations in spectral measurements. This suggests that it gives an indication of the optical properties of a sample, however it may not be precise. 

When modelling the measured tissue phantom diffuse reflectance spectra using the forward models, Yudovsky performs best followed closely by Jacques, with Modified Beer-Lambert performing significantly worse. The $NRMSE$ calculated for relative data is an improvement on those calculated for quantitative data, suggesting that these models are reproducing the shape of the spectra well even when relatively constant offsets are present. When introducing a third dye ($CV$) the absorbance peak of this dye is noticeably wavelength shifted in experimental measurements compared to the theoretical spectra which then appears distorted. This shift appears to change with varying concentrations of intralipid suggesting some interaction there. 

When fitting the models to experimental data for parameter recovery, Yudovsky and Jacques produce excellent parameter recovery for $AR1$ and $AR14$ in both quantitative and relative regimes. For $c_{tot}$ and $I$ no models were able to recover the parameters well, whereas all parameters were recovered well by Yudovsky and Jacques when fitted to Monte Carlo data. This is suggestive of some overlap in the effects of $c_{tot}$ and $I$ on the spectra. In the three-dye configuration $CV$ is recovered reasonably, despite the peak being outside of the region considered for fitting. The shift in the $CV$ peak, however, distorts the rest of the spectrum leading to very poor recovery of all other parameters. This demonstrates the importance of prior understanding of the chromophores for any model to recover parameters accurately. 

This work is associated with some limitations.
Firstly, whilst every effort was made to synthesise optical phantoms resembling biological tissue, measurements of true biological tissue could not be used for this work due to a lack of reliable ground truth parameters in tissue.
Similarly, haemoglobin chromophores could not be utilised in the phantoms due to their oxygen sensitivity which could not be controlled within the spectrophotometer measurement set-up. This led us to rely on stable synthetic dyes instead.
Secondly, the shift in absorbance of the third chromophore made it difficult to model thereby limiting the investigation of the impact of additional chromophores.
Thirdly, whilst the IAD method produces results similar to our ground truth, there remain some differences between IAD output and the expected ground truth.
This could impact the quality of the intralipid scattering model used in this work.
Fourthly, these models assume a planar surface for measurement, however the surface may deviate from this. In these cases, Spatial Frequency Domain methods may be an alternative to obtain accurate $\mu_a$ and $\mu_s'$ values for tissues\cite{Gioux2019}. 
Finally, the precise measurement of extinction coefficients of haemoglobin in biological tissue is not possible due to the scattering nature of these media.
However, it is shown in our work that the medium chosen for the dye can have an impact on the extinction coefficients and therefore the quality of the models. For this reason, the optical properties of tissue must be well defined to mimic the quality of results in this work.

Future work should include investigation of these models for use with hyperspectral imaging cameras which may not have as densely sampled spectra as these have been shown to be likely methods of obtaining these diffuse reflectance spectra intra-operatively. \cite{Clancy2020}. There could also be greater investigation of including further chromophores whose absorbance behaviour is well modelled and optimisation of the fitting region. Finally, two-layer models should also be considered for tissue structures which exist in layered configurations, such as skin.  

In conclusion, the Yudovsky single layer model works well for modelling of tissue that can be approximated by a semi-infinite, homogeneous slab. Jacques is also able to well approximate this with a simpler model. All models require appropriate prior knowledge of the $mu_a$ and $mu_s'$ properties of the tissues to allow for a good quality of parameter recovery.

\section*{Acknowledgements}
The authors would like to thank Jenasee Mynerich of King's College London who provided the CT imaging for the tissue phantom work presented.

This study/project is funded by the NIHR [NIHR202114]. The views expressed are those of the author(s) and not necessarily those of the NIHR or the Department of Health and Social Care. 
This work was supported by core funding from the Wellcome/EPSRC [WT203148/Z/16/Z; NS/A000049/1].
TV is supported by a Medtronic / RAEng Research Chair [RCSRF1819\textbackslash7\textbackslash34].

TV and JS are co-founders and shareholders of Hypervision Surgical Ltd, London, UK.

For the purpose of open access, the authors have applied a CC BY public copyright licence to any Author Accepted Manuscript version arising from this submission.

\bibliography{Paper2}

\appendix
\begin{table}[htb!]
    \centering
    \caption{Table displaying single layer model hyperparameters fitted to Monte Carlo datasets of each refractive index (black) with available literature values (blue).}
    \begin{tabular}{|cc|cc|c|cc|}
        \hline
        \multirow{2}{*}{Model} & \multirow{2}{*}{Parameter} & \multicolumn{5}{c|}{Refractive index} \\
         & & \multicolumn{2}{c|}{1.33} & 1.35 & \multicolumn{2}{c|}{1.44} \\
        \hline
        \multirow{6}{*}{Yudovsky 2009} & $M_1$ & \multicolumn{2}{c|}{-0.0253} & -0.0257 & -0.0254 & \textcolor{blue}{-0.0247} \\
        & $M_2$ & \multicolumn{2}{c|}{0.0166} & 0.0159 & 0.0135 & \textcolor{blue}{0.0137} \\
        & $M_3$ & \multicolumn{2}{c|}{2.873} & 2.873 & 2.873 & \textcolor{blue}{2.873} \\
        & $M_4$ & \multicolumn{2}{c|}{1.64} & 1.64 & 1.64 & \textcolor{blue}{1.64} \\
        & $M_5$ & \multicolumn{2}{c|}{0.0123} & 0.0124 & 0.0120 & \textcolor{blue}{0.0116} \\
        & $M_6$ & \multicolumn{2}{c|}{1.02} & 1.02 & 1.02 & \textcolor{blue}{1.02} \\
        \hline 
        \multirow{3}{*}{Jacques 1999} & $M_1$ & 7.0188 & \textcolor{blue}{6.3744} & 7.1185 & \multicolumn{2}{c|}{7.0438} \\
        & $M_2$ & 0.2464 & \textcolor{blue}{0.35688} & 0.2750 & \multicolumn{2}{c|}{0.6902} \\
        & $M_3$ & 4.2241 & \textcolor{blue}{3.4739} & 4.2571 & \multicolumn{2}{c|}{4.1449} \\
        \hline
        \multirow{3}{*}{Modified Beer-Lambert} & $M_1$ & \multicolumn{2}{c|}{0.283} & 0.308 & \multicolumn{2}{c|}{0.256} \\
        & $M_2$ & \multicolumn{2}{c|}{0.009} & 0.008 & \multicolumn{2}{c|}{0.014} \\
        & $M_3$ & \multicolumn{2}{c|}{0.203} & 0.311 & \multicolumn{2}{c|}{0.274} \\
        \hline
    \end{tabular}
    \label{tb:fittedmodelparams}
\end{table}
\FloatBarrier

\begin{figure}[htb!]
    \centering
    \includegraphics[width=0.65\textwidth]{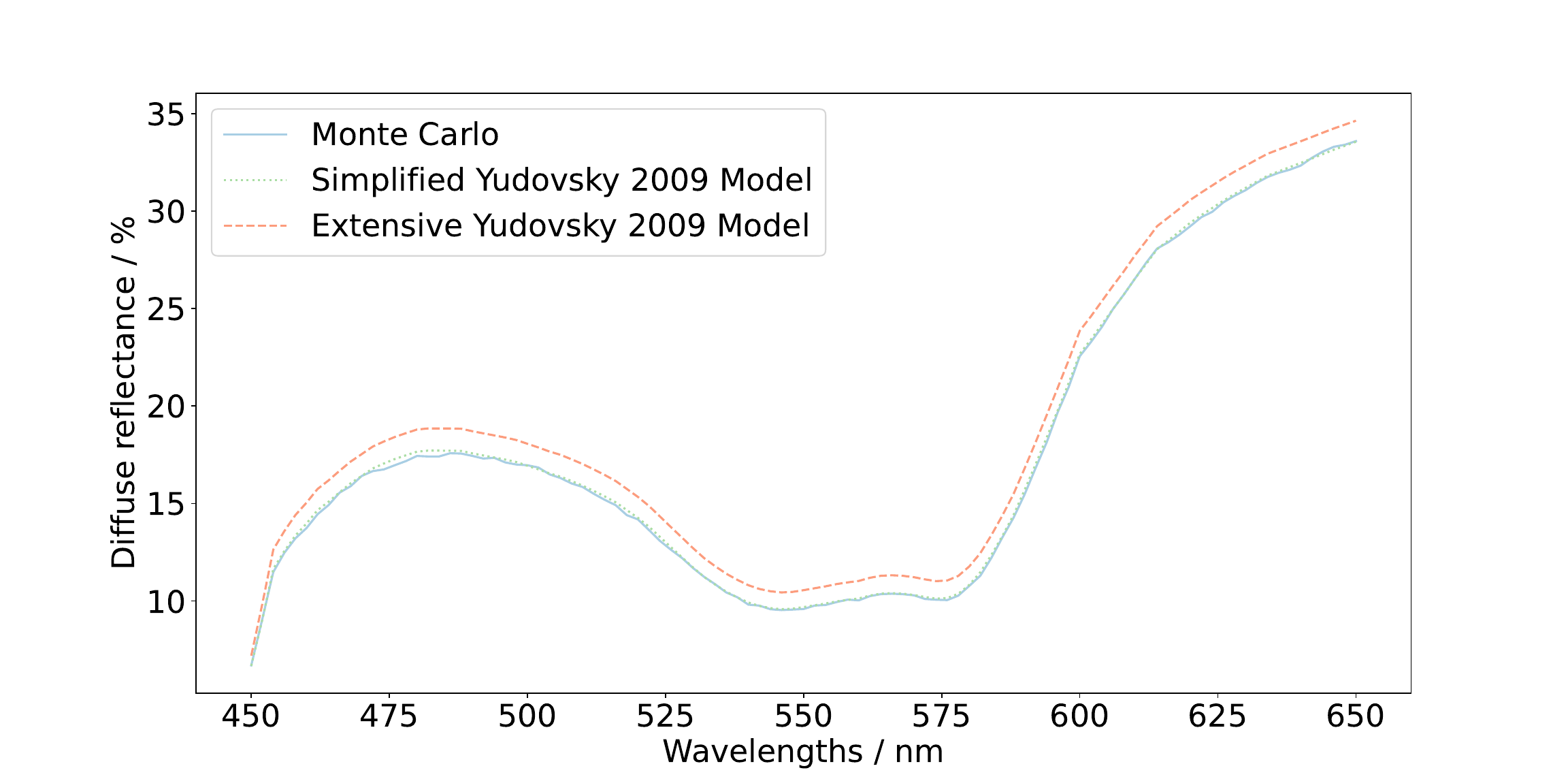}
    \caption{Example of difference in forwards model similarity to Monte Carlo simulation (blue solid) between the literature extensive Yudovsky 2009 (red dashed) and simplified Yudovsky 2009 (green dotted) models using ground truth variables for a refractive index of 1.44.}
 \label{fig:badYudovsky}
\end{figure}
\FloatBarrier


\begin{table}[htb!]
    \centering
    \caption{The gradient $m$, offset $c$, Pearson $r$, and $p$ values of the linear regression line between the fitted tissue parameters and their ground truth displayed with their median (inter-quartile range) absolute percentage errors. This is shown for each variable and for each refractive index dataset when extracted by fitting Yudovsky 2009 (Y), Jacques 1999 (J), or Modified Beer-Lambert (BL) to the Monte-Carlo dataset. All presented to 3s.f.}
    \begin{tabular}{|ccc|ccccc|}
        \hline
        parameter & model & refractive & $m$ & $c$ & $r$ & $p$ & median (inter-quartile range) \\
        & & index & (ideal =1) & (ideal = 0) & (ideal = 1) & (ideal = 0) & absolute percentage error (\%)\\
        \hline
        \multirow{9}{*}{$StO_2$} & \multirow{3}{*}{Y} & 1.33 & 1.00 & -0.373 & 1.00 & 2.39$\times 10^{-150}$ & 1.23 (3.01) \\
        & & 1.35 & 0.999 & 0.0672 & 1.00 & 4.93$\times 10^{-149}$ & 0.815 (2.10) \\
        & & 1.44 & 0.999 & -0.074 & 1.00 & 4.87$\times 10^{-153}$ & 0.913 (1.92) \\
        \cline{2-8}
        & \multirow{3}{*}{J} & 1.33 & 1.00 & 1.733 & 0.979 & 1.44$\times 10^{-69}$ & 3.40 (9.43) \\
        & & 1.35 & 1.03 & 1.03 & 0.980 & 2.18$\times 10^{-70}$ & 3.97 (12.9) \\
        & & 1.44 &  1.04 & 0.207 & 0.986 & 1.64$\times 10^{-77}$ & 2.21 (4.97) \\
        \cline{2-8}
        & \multirow{3}{*}{BL} & 1.33 & 0.712 & 1.56 & 0.855 & 1.17$\times 10^{-29}$ & 39.4 (38.5) \\
        & & 1.35 & 0.679 & 3.92 & 0.838 & 3.87$\times 10^{-30}$ & 33.3 (31.4) \\
        & & 1.44 & 0.766 & -2.73 & 0.838 & 1.70$\times 10^{-27}$ & 43.0 (49.8) \\
        \hline
        \multirow{9}{*}{$f_{blood}$} & \multirow{3}{*}{Y} & 1.33 & 0.976 & -0.0206 & 0.980 & 2.76$\times 10^{-70}$ & 5.74 (6.19) \\
        & & 1.35 & 0.962 & 0.0365 & 0.975 & 8.29$\times 10^{-66}$ & 4.58 (6.98) \\
        & & 1.44 & 0.927 & 0.152 & 0.982 & 3.60$\times 10^{-73}$ & 5.68 (6.08) \\
        \cline{2-8}
        & \multirow{3}{*}{J} & 1.33 & 1.17 & 0.187 & 0.915 & 2.21$\times 10^{-40}$ & 9.51 (28.7) \\
        & & 1.35 & 1.16 & 0.167 & 0.912 & 9.18$\times 10^{-40}$ & 11.0 (20.1) \\
        & & 1.44 & 1.15 & 0.141 & 0.928 & 1.10$\times 10^{-43}$ & 7.26 (16.2) \\
        \cline{2-8}
        & \multirow{3}{*}{BL} & 1.33 & 0.487 & 0.431 & 0.641 & 7.16$\times 10^{-13}$ & 46.4 (28.2) \\
        & & 1.35 & 0.461 & 0.478 & 0.575 & 3.99$\times 10^{-10}$ & 45.8 (41.8) \\
        & & 1.44 & 0.339 & 0.652 & 0.582 & 2.05$\times 10^{-10}$ & 49.6 (32.2) \\
        \hline
        \multirow{9}{*}{$a$} & \multirow{3}{*}{Y} & 1.33 & 0.948 & 0.145 & 0.993 & 2.58$\times 10^{-93}$ & 5.06 (5.92) \\
        & & 1.35 & 0.968 & -0.192 & 0.995 & 4.71$\times 10^{-102}$ & 4.05 (5.57) \\
        & & 1.44 & 1.05 & -2.59 & 0.992 & 1.67$\times 10^{-89}$ & 3.90 (4.73) \\
        \cline{2-8}
        & \multirow{3}{*}{J} & 1.33 & 0.909 & 7.23 & 0.952 & 4.13$\times 10^{-52}$ & 7.09 (22.0) \\
        & & 1.35 & 0.929 & 5.72 & 0.966 & 3.04$\times 10^{-59}$ & 9.16 (16.2) \\
        & & 1.44 & 0.934 & 5.88 & 0.959 & 1.05$\times 10^{-55}$ & 4.54 (15.3) \\
        \cline{2-8}
        & \multirow{3}{*}{BL} & 1.33 & -0.197 & 74.7 & -0.554 & 2.23$\times 10^{-9}$ & 63.8 (133) \\
        & & 1.35 & -0.142 & 73.1 & -0.448 & 2.89$\times 10^{-6}$ & 103 (206) \\
        & & 1.44 & -0.488 & 78.1 & -0.706 & 2.28$\times 10^{-16}$ & 40.0 (114) \\
        \hline
        \multirow{9}{*}{$b$} & \multirow{3}{*}{Y} & 1.33 & 0.978 & 0.0663 & 0.998 & 3.53$\times 10^{-121}$ & 1.38 (2.47) \\
        & & 1.35 & 0.989 & 0.0314 & 0.998 & 5.55$\times 10^{-125}$ & 1.82 (2.85) \\
        & & 1.44 & 0.991 & 0.0185 & 0.999 & 1.33$\times 10^{-140}$ & 1.50 (2.92) \\
        \cline{2-8}
        & \multirow{3}{*}{J} & 1.33 & 0.977 & 0.216 & 0.910 & 3.76$\times 10^{-39}$ & 3.45 (16.0) \\
        & & 1.35 & 1.01 & 0.212 & 0.906 & 2.35$\times 10^{-38}$ & 5.29 (25.2) \\
        & & 1.44 & 1.01 & 0.120 & 0.963 & 1.15$\times 10^{-57}$ & 2.86 (9.11) \\
        \cline{2-8}
        & \multirow{3}{*}{BL} & 1.33 & -0.104 & 0.360 & -0.413 & 1.92$\times 10^{-5}$ & 95.2 (4.85) \\
        & & 1.35 & -0.0888 & 0.303 & -0.394 & 5.02$\times 10^{-5}$ & 94.1 (5.60) \\
        & & 1.44 & -0.0859 & 0.327 & -0.446 & 3.38$\times 10^{-6}$ & 95.4 (5.68) \\
        \hline
    \end{tabular}    \label{tb:singleparamtrendsfull}
\end{table}
\FloatBarrier

\begin{figure}[htb!]
    \centering
    \includegraphics[width=0.7\textwidth]{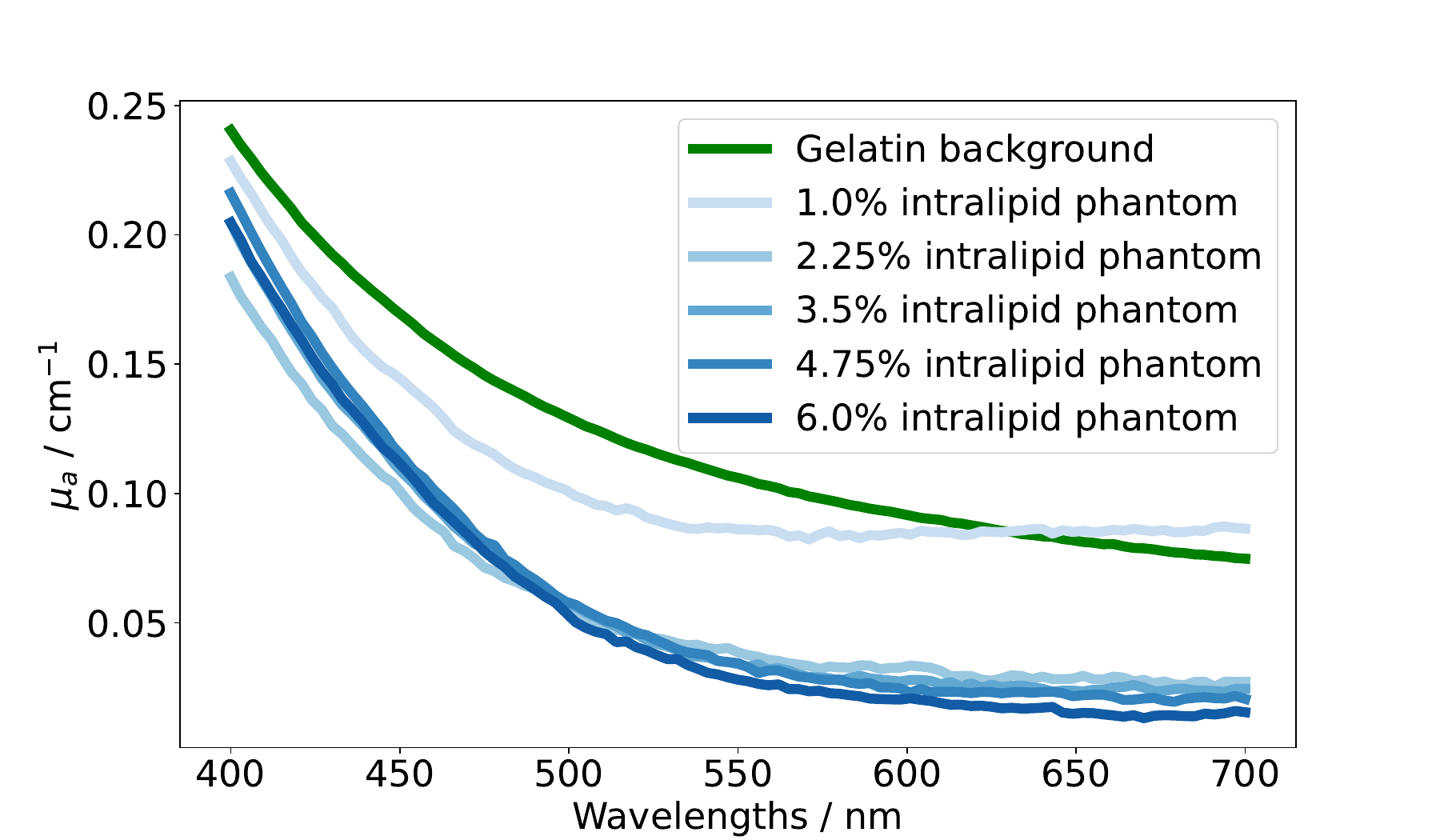}
    \caption{Background $\mu_a$ using absorption of pure gelatin solution compared to the IAD returned $\mu_a$ of the phantoms containing intralipid but no dyes.}
 \label{fig:muaback}
\end{figure}
\FloatBarrier

\begin{figure}[h!]
    \centering
    \includegraphics[width=0.7\textwidth]{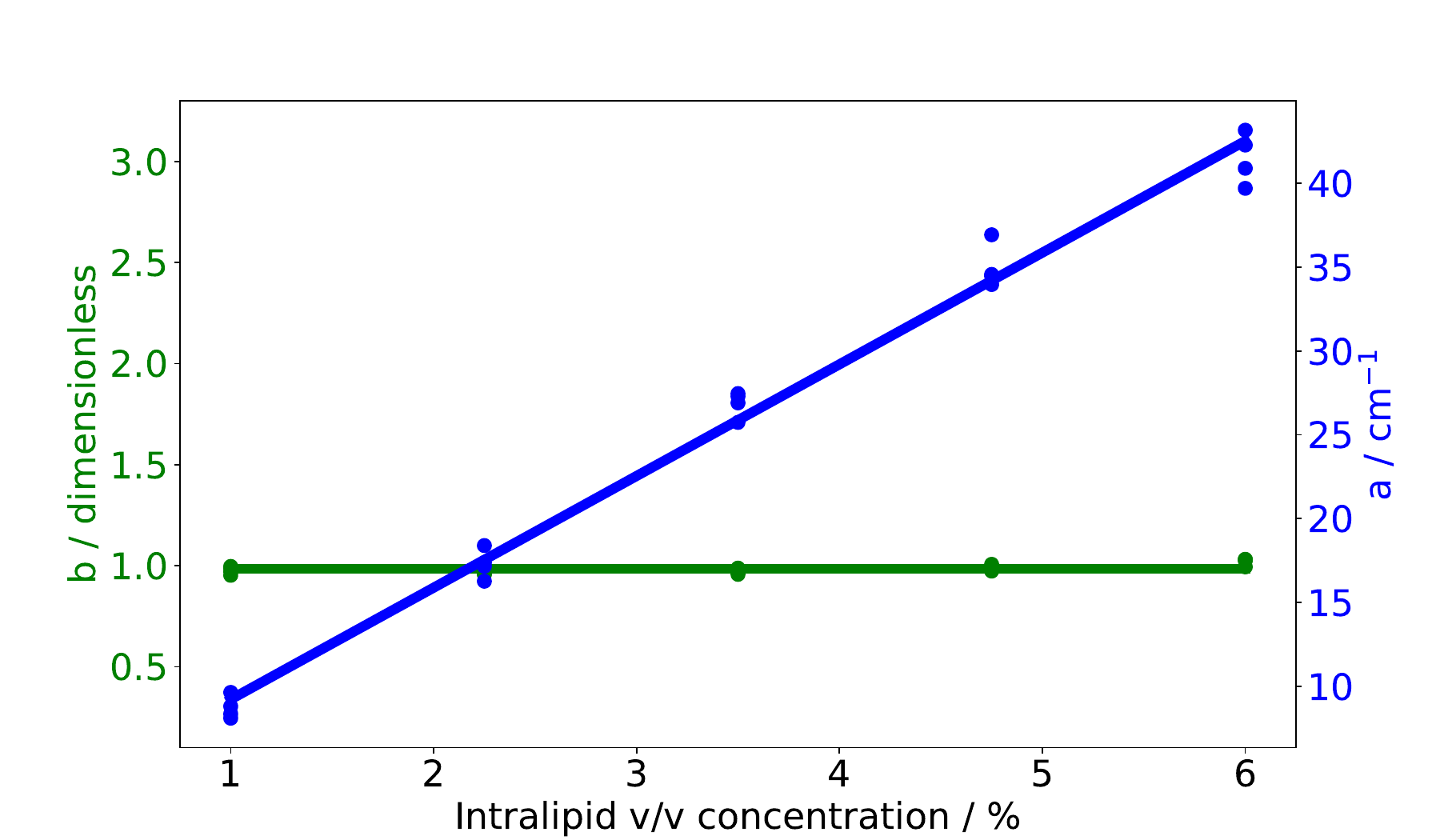}
    \caption{Trend in $a$ with intralipid concentration alongside linear regression. The trend in $b$ with intralipid concentration is also plotted alongside the median $b$ value.}
 \label{fig:atrend}
\end{figure}
\FloatBarrier


\begin{table}[htb!]
    \centering
    \caption{The regression gradient $m$, offset $c$, Pearson $r$, and $p$ values, and the median (inter-quartile range) absolute percentage errors for each variable when extracted by fitting Yudovsky 2009 (Y), Jacques 1999 (J), or Modified Beer-Lambert (BL) to measured tissue phantom spectra for both quantitative and relative data. All presented to 3s.f.}
    \begin{tabular}{|ccc|ccccc|}
        \hline
        parameter & model & Quantitative (Q) & $m$ & $c$ & $r$ & $p$ & median (inter-quartile range) \\
        & & or Relative (R) & (ideal =1) & (ideal = 0) & (ideal = 1) & (ideal = 0) & absolute percentage error (\%)\\
        \hline
        \multirow{6}{*}{$AR1$} & \multirow{2}{*}{Y} & Q & 1.00 & 0.0132 & 0.997 & 5.75$\times 10^{-38}$ & 1.59 (11.0) \\
        & & R & 1.01 & -0.0166 & 0.998 & 6.72$\times 10^{-40}$ & 4.42 (8.31) \\
        \cline{2-8}
        & \multirow{2}{*}{J} & Q & 0.976 & 0.0481 & 0.989 & 6.16$\times 10^{-29}$ & 7.02 (20.2) \\
        & & R & 0.905 & 0.0442 & 0.934 & 2.67$\times 10^{-16}$ & 10.4 (23.7) \\
        \cline{2-8}
        & \multirow{2}{*}{BL} & Q & 0.850 & -0.148 & 0.756 & 1.50$\times 10^{-7}$ & 83.5 (57.0) \\
        & & R & 0.730 & -0.107 & 0.939 & 7.84$\times 10^{-17}$ & 50.0 (31.4)\\
        \hline
        \multirow{6}{*}{$AR14$} & \multirow{2}{*}{Y} & Q & 1.00 & -0.0140 & 0.997 & 5.75$\times 10^{-38}$ & 1.36 (8.69) \\
        & & R & 1.01 & 0.00675 & 0.998 & 6.72$\times 10^{-40}$ & 2.48 (5.76) \\
        \cline{2-8}
        & \multirow{2}{*}{J} & Q & 0.976 & -0.0244 & 0.989 & 6.16$\times 10^{-29}$ & 7.02 (16.3) \\
        & & R & 0.905 & 0.0508 & 0.934 & 2.67$\times 10^{-16}$ & 6.34 (10.0) \\
        \cline{2-8}
        & \multirow{2}{*}{BL} & Q & 0.850 & 0.297 & 0.756 & 1.50$\times 10^{-7}$ & 80.4 (72.1) \\
        & & R & 0.730 & 0.377 & 0.939 & 7.84$\times 10^{-17}$ & 37.5 (66.7) \\
        \hline
        \multirow{6}{*}{$I$} & \multirow{2}{*}{Y} & Q & 2.41 & 1.09 & 0.648 & 2.61$\times 10^{-5}$ & 125 (276) \\
        & & R & 0.998 & 1.37 & 0.310 & 6.96$\times 10^{-2}$ & 93.9 (30.0) \\
        \cline{2-8}
        & \multirow{2}{*}{J} & Q & 1.62 & 3.73 & 0.540 & 8.11$\times 10^{-4}$ & 122 (288) \\
        & & R & -0.767 & 4.76 & -0.451 & 6.61$\times 10^{-3}$ & 100 (55.7) \\
        \cline{2-8}
        & \multirow{2}{*}{BL} & Q & -1.75 & 11.8 & -0.705 & 2.29$\times 10^{-6}$ & 100 (172) \\
        & & R & -0.484 & 3.94 & -0.157 & 3.66$\times 10^{-1}$ & 100 (0.00) \\
        \hline
        \multirow{6}{*}{$c_{tot}$} & \multirow{2}{*}{Y} & Q & 1.46 & 9.25 & 0.500 & 2.23$\times 10^{-3}$ & 102 (226) \\
        & & R & 0.917 & 4.85 & 0.303 & 7.72$\times 10^{-2}$ & 66.1 (46.8) \\
        \cline{2-8}
        & \multirow{2}{*}{J} & Q & 2.00 & 6.76 & 0.684 & 5.95$\times 10^{-6}$ & 100 (260) \\
        & & R & -0.265 & 27.8 & -0.0959 & 5.84$\times 10^{-1}$ & 89.7 (267) \\
        \cline{2-8}
        & \multirow{2}{*}{BL} & Q & 0.356 & 3.02 & 0.721 & 1.01$\times 10^{-6}$ & 35.2 (21.6) \\
        & & R & 0.309 & 3.23 & 0.689 & 4.63$\times 10^{-6}$ & 44.6 (23.6) \\
        \hline
    \end{tabular}
    \label{tb:phantomparamsfull}
\end{table}
\FloatBarrier

\begin{figure}[htb!]
    \centering
    \begin{subfigure}{0.8\textwidth}
        \includegraphics[width=\textwidth]{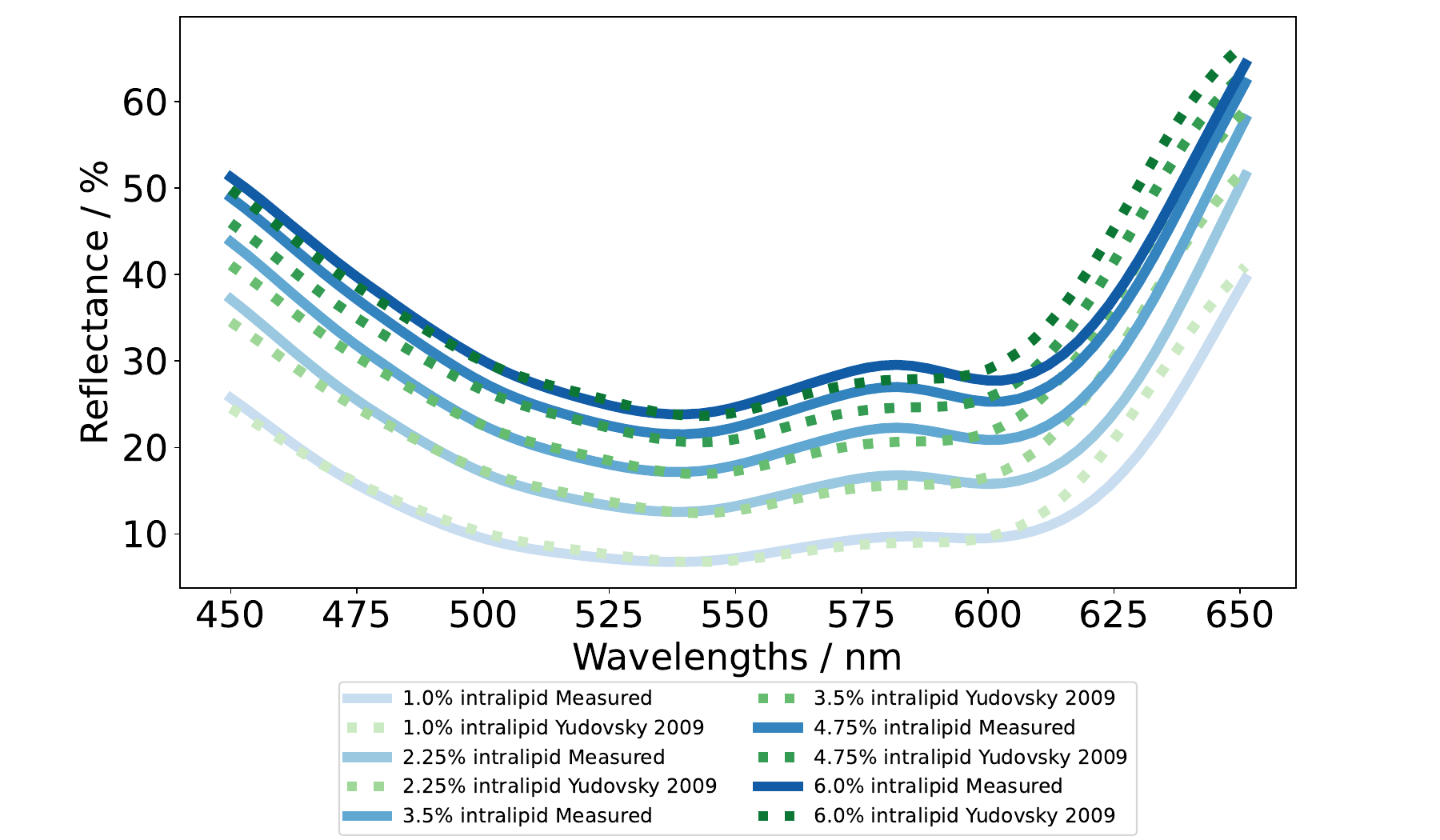}
        \caption{}
        \label{fig:3phantomforwardsquant}
    \end{subfigure}
    \begin{subfigure}{0.8\textwidth}
        \includegraphics[width=\textwidth]{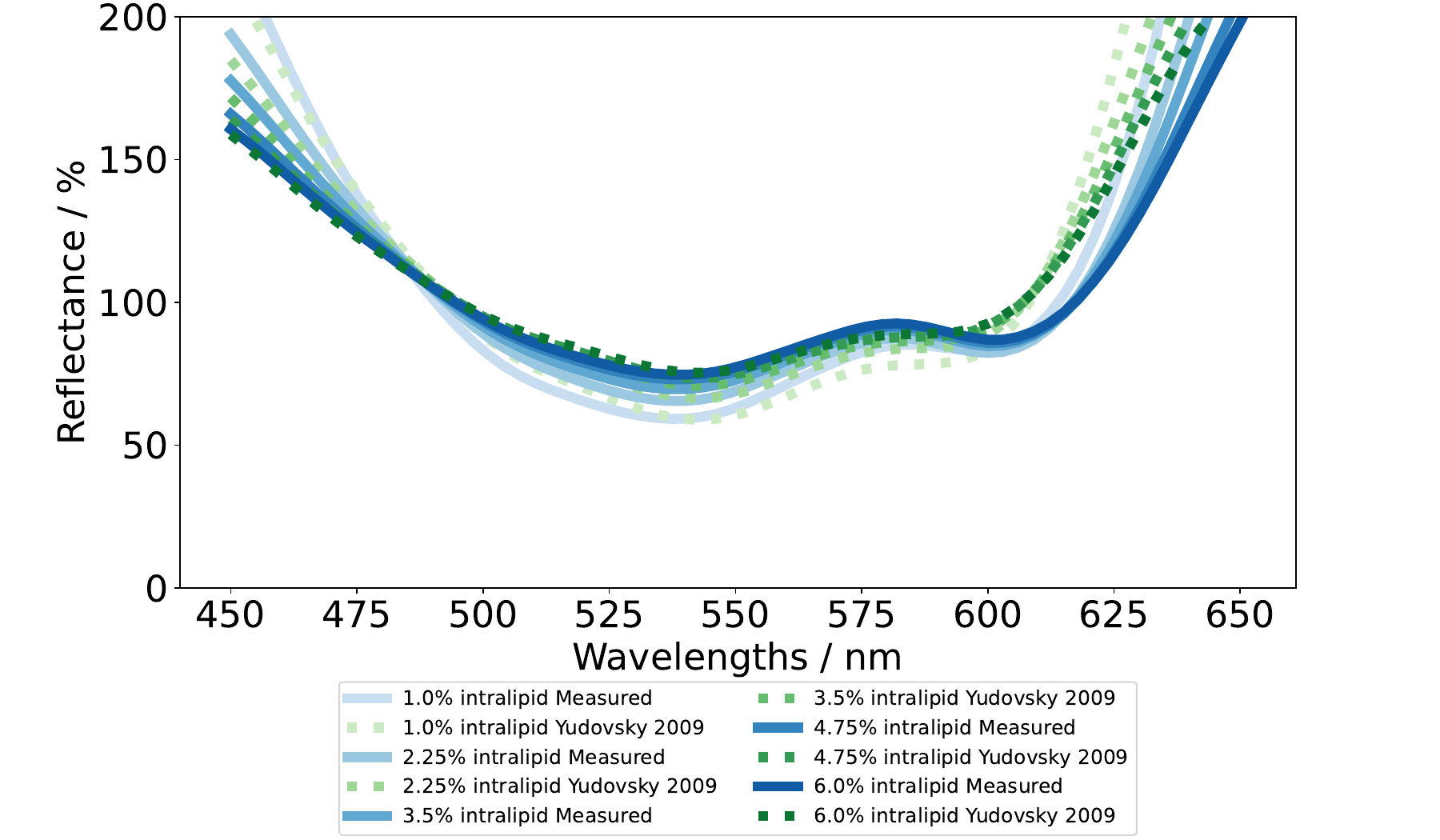}
        \caption{}
        \label{fig:3phantomforwardsnorm}
    \end{subfigure}
    \caption{Examples of Measured (blue) spectrum for one 3-dye configuration at a variety of intralipid concentrations compared to the respective spectra generated from the Yudovsky 2009 (green dotted) model using the ground truth parameters to generate quantitative (\ref{fig:3phantomforwardsquant}) or relative (\ref{fig:3phantomforwardsnorm}) spectra.}
 \label{fig:3phantomforwards}
\end{figure}
\FloatBarrier

\begin{table}[htb!]
    \centering
    \caption{The median (inter-quartile range) absolute percentage errors for each variable when extracted by fitting Yudovsky 2009 (Y), Jacques 1999 (J), or Modified Beer-Lambert (BL) to measured 3-dye tissue phantom spectra for both quantitative and relative data. All presented to 3s.f.}
    \begin{tabular}{|ccc|c|}
        \hline
        parameter & model & Quantitative (Q) & median (inter-quartile range) \\
        & & or Relative (R) & absolute percentage error (\%)\\
        \hline
        \multirow{6}{*}{$AR1$} & \multirow{2}{*}{Y} & Q & 22.7 (25.8) \\
        & & R & 14.9 (9.14) \\
        \cline{2-4}
        & \multirow{2}{*}{J} & Q & 30.6 (29.8) \\
        & & R & 49.1 (84.6) \\
        \cline{2-4}
        & \multirow{2}{*}{BL} & Q & 86.7 (58.3) \\
        & & R & 77.1 (63.5) \\
        \hline
        \multirow{6}{*}{$AR14$} & \multirow{2}{*}{Y} & Q & 25.3 (23.0) \\
        & & R & 40.4 (23.7) \\
        \cline{2-4}
        & \multirow{2}{*}{J} & Q & 28.0 (28.8) \\
        & & R & 97.3 (219) \\
        \cline{2-4}
        & \multirow{2}{*}{BL} & Q & 129 (74.7) \\
        & & R & 122 (113) \\
        \hline
        \multirow{6}{*}{$CV$} & \multirow{2}{*}{Y} & Q & 13.3 (20.7) \\
        & & R & 19.2 (16.1) \\
        \cline{2-4}
        & \multirow{2}{*}{J} & Q & 9.25 (17.0) \\
        & & R & 21.0 (79.4) \\
        \cline{2-4}
        & \multirow{2}{*}{BL} & Q & 19.0 (13.2) \\
        & & R & 17.4 (14.3) \\
        \hline
        \multirow{6}{*}{$I$} & \multirow{2}{*}{Y} & Q & 145 (285) \\
        & & R & 88.4 (17.9) \\
        \cline{2-4}
        & \multirow{2}{*}{J} & Q & 131 (231) \\
        & & R & 100 (11.2) \\
        \cline{2-4}
        & \multirow{2}{*}{BL} & Q & 100 (67.1) \\
        & & R & 100 (2.53 $\times 10^{-10}$) \\
        \hline
        \multirow{6}{*}{$c_{tot}$} & \multirow{2}{*}{Y} & Q & 100 (285) \\
        & & R & 63.2 (24.1) \\
        \cline{2-4}
        & \multirow{2}{*}{J} & Q & 100 (206) \\
        & & R & 81.0 (3510) \\
        \cline{2-4}
        & \multirow{2}{*}{BL} & Q & 36.3 (36.9) \\
        & & R & 45.7 (66.9) \\
        \hline
    \end{tabular}
    \label{tb:3phantomparams}
\end{table}
\FloatBarrier
\section{IAD validation}\label{ap:IAD}
\begin{figure}[ht!]
    \begin{subfigure}{0.49\textwidth}
        \includegraphics[width=\textwidth]{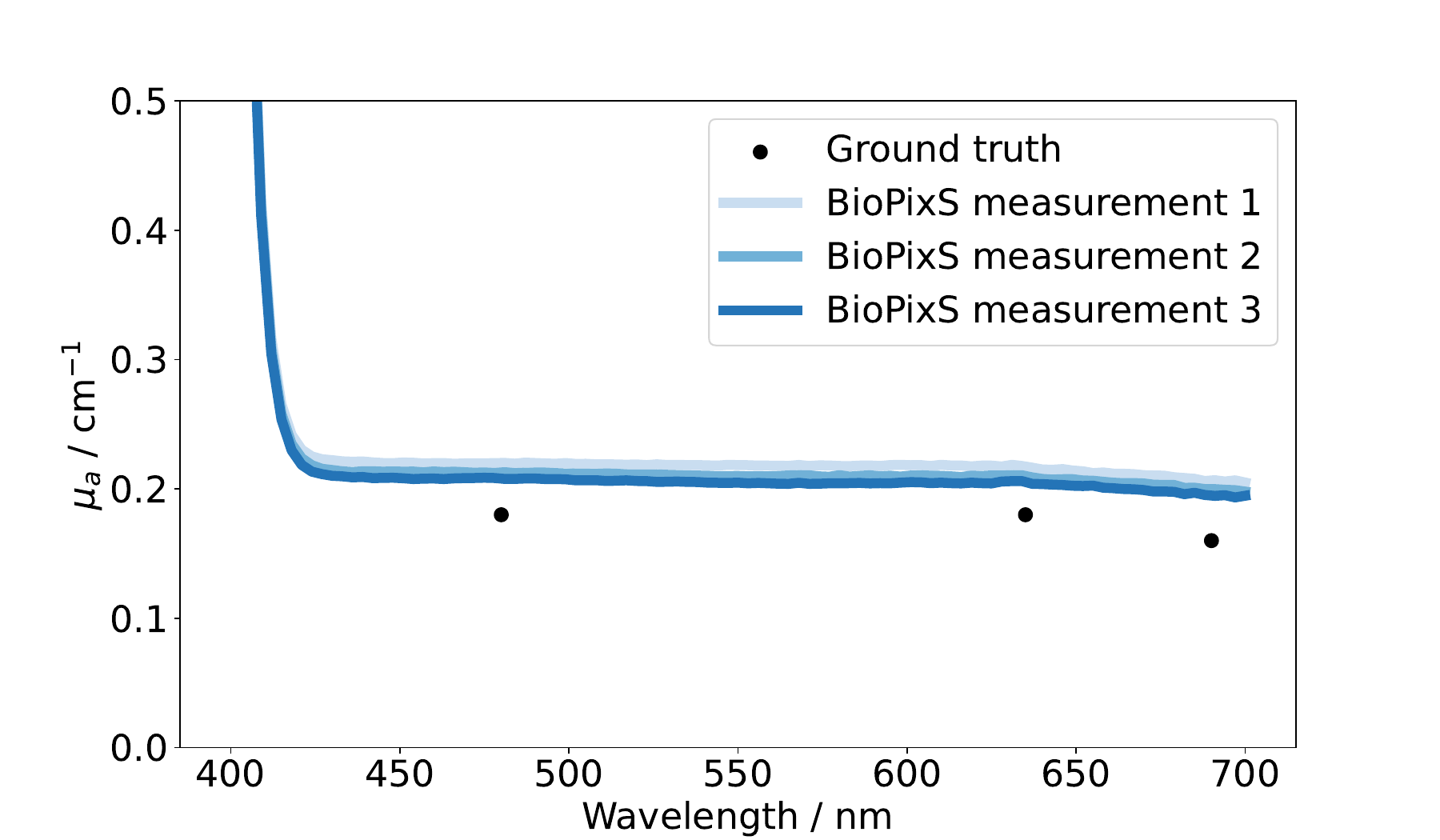}
        \caption{}
        \label{fig:BioPixSmua}
    \end{subfigure}
    \begin{subfigure}{0.49\textwidth}
        \includegraphics[width=\textwidth]{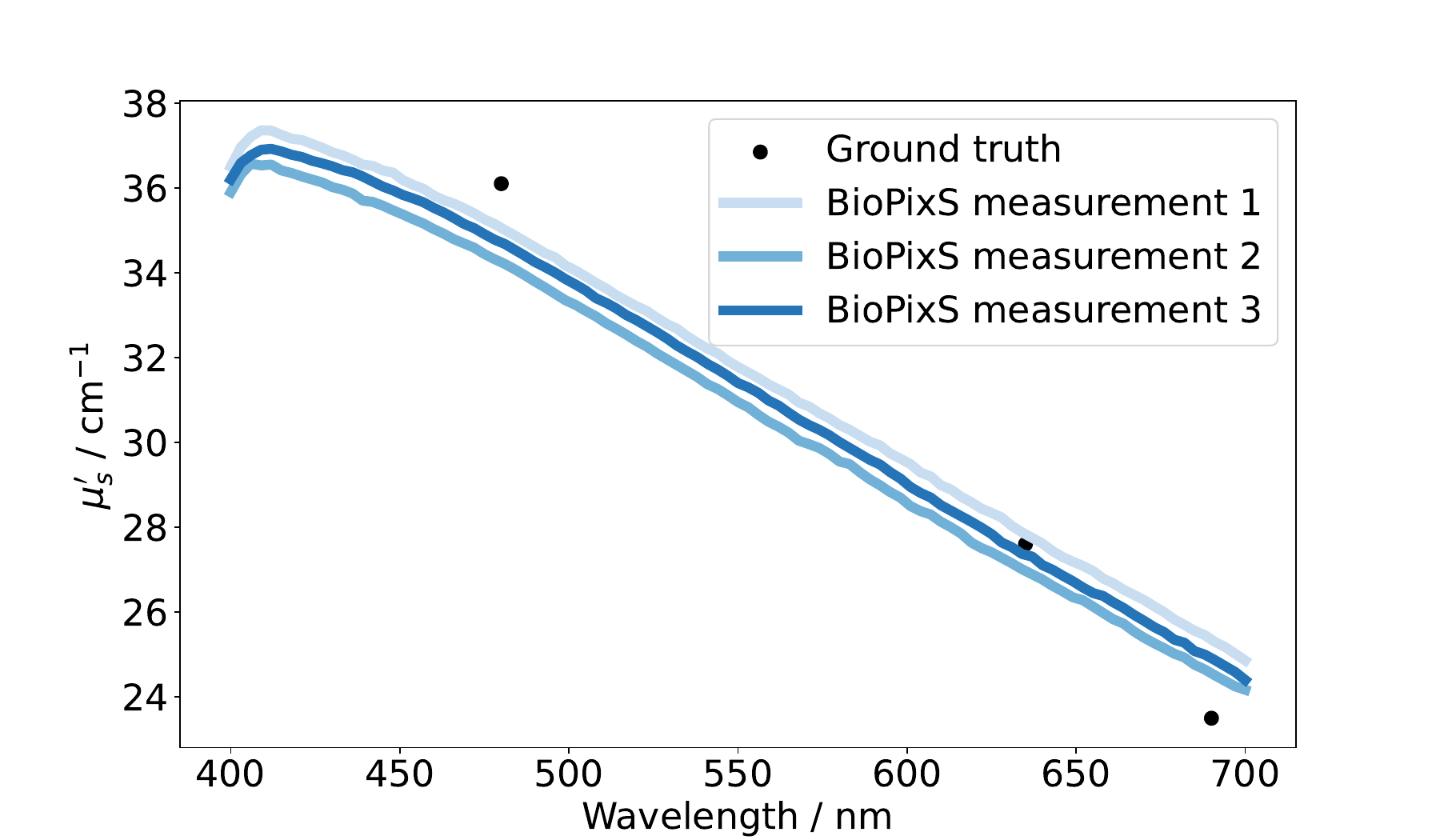}
        \caption{}
        \label{fig:BioPixSmusp}
    \end{subfigure}
    Figure \ref{fig:BioPixSmua} shows the IAD outputted $\mu_a$ and Figure \ref{fig:BioPixSmusp} shows the IAD outputted $\mu_s'$ compared to the well-characterised ground truth values for a BioPixS optical phantom measured on three separate days.
\end{figure}

To demonstrate that optical properties output by IAD are reproducible and accurate, the output properties from IAD analysis of measurements of the same BioPixS phantom on three days are shown in Figure~\ref{fig:BioPixSmua} and Figure~\ref{fig:BioPixSmusp}. Whilst there is some variability, the results show that the optical properties returned by this method accurately reflect the ground truth properties.

\end{document}